\documentclass[fleqn,usenatbib]{mnras}

\usepackage{newtxtext,newtxmath}
\usepackage[T1]{fontenc}

\DeclareRobustCommand{\VAN}[3]{#2}
\let\VANthebibliography\thebibliography
\def\thebibliography{\DeclareRobustCommand{\VAN}[3]{##3}\VANthebibliography}

\usepackage{enumitem}

\usepackage{makecell}
\usepackage{physics}
\usepackage{mathtools,lipsum}

\usepackage{cuted}
\usepackage{multirow}

\usepackage{float}
\usepackage{graphicx}	% Including figure files
\usepackage{amsmath}	% Advanced maths commands
\usepackage[dvipsnames]{xcolor}

\def\p{\partial}
\def\fnl{f_\text{NL}}
\def\LCDM{\Lambda\text{CDM}}
\usepackage{multirow}	% per cm-squared

\title[Next-generation spectroscopic survey forecasts]{Cosmological Fisher
forecasts for next-generation spectroscopic surveys}

\author[W. d'Assignies et al.  ]{
William d'Assignies D.$^{1,2,3}$\thanks{E-mail: wdoumerg@ifae.es}, Cheng Zhao$^{1}$\thanks{E-mail: cheng.zhao@epfl.ch}, Jiaxi Yu$^{1}$ 
and Jean-Paul Kneib$^{1,4}$
\\
$^{1}$ Laboratory of Astrophysics, École Polytechnique Fédérale de Lausanne (EPFL), Observatoire de Sauverny, CH-1290 Versoix, Switzerland.\\
$^{2}$ Institut de Física d'Altes Energies (IFAE), The Barcelona Institute of Science and Technology, Campus UAB, 08193 Bellaterra (Barcelona) Spain.\\
$^{3}$ Physics institute of the École Normale Supérieure PSL, 24 rue Lhomond, 75005 Paris, France.\\
$^{4}$ Aix Marseille Université, CNRS, LAM (Laboratoire d’Astrophysique de Marseille) UMR 7326, F13388, Marseille, France
}

\date{Accepted XXX. Received YYY; in original form ZZZ}

\pubyear{2022}

\begin{document}
\label{firstpage}
\pagerange{\pageref{firstpage}--\pageref{lastpage}}
\maketitle

\begin{abstract}
Next-generation spectroscopic surveys such as the MegaMapper, MUltiplexed Survey Telescope (MUST), MaunaKea Spectroscopic Explorer (MSE), and WideField Spectroscopic Telescope (WST) are foreseen to increase the number of galaxy/quasar redshifts by an order of magnitude, with hundred millions of spectra that will be measured at $z>2$.  
We perform a Fisher matrix analysis for these surveys on the baryonic acoustic oscillation (BAO), the redshift-space distortion (RSD) measurement, the non-Gaussianity amplitude $\fnl$, and the total neutrino mass $M_\nu$. For BAO and RSD  parameters, these surveys may achieve precision at sub-per-cent level (<0.5 per cent), representing an improvement of factor 10 w.r.t. the latest database. For NG, these surveys may reach an accuracy of $\sigma(\fnl)\sim 1$. They can also put a tight constraint on $M_\nu$ with $\sigma(M_\nu) \sim 0.02\,\rm eV$ if we do joint analysis with Planck and even $ 0.01\,\rm eV$ if combined with other data.
In addition, we introduce a general survey model, to derive the cosmic volume and number density of tracers, given instrumental facilities and survey strategy. Using our Fisher formalism, we can explore (continuously) a wide range of survey observational parameters, and propose different survey strategies that optimise the cosmological constraints.
Fixing the fibre number and survey duration, we show that the best strategy for $\fnl$ and $M_\nu$ measurement is to observe large volumes, despite the noise increase. However, the strategy differs for the apparent magnitude limit. Finally, we prove that increasing the fibre number improves $M_{\nu}$ measurement but not significantly $\fnl$.

\end{abstract}
 

%%%%%%%%%%%%%%%%%%%%%%%%%%%%%%%%%%%%%%%%%%%%%%%%%%

%%%%%%%%%%%%%%%%% BODY OF PAPER %%%%%%%%%%%%%%%%%%

\section{Introduction}
%blabla neutrino, blabla hubble tension, blabla bao...

Massive high redshift spectroscopic survey aims at exploring baryon acoustic oscillations
(BAO) and the growth of structure through redshift-space distortions (RSD) with large-scale structures (LSS) in the Universe, by probing the 3D distribution of galaxies and quasars in a wide area.  LSS also provides one of our best windows on fundamental
physics, such as properties of the early Universe with the non-Gaussian primordial fluctuations, or the sum of neutrino mass (NM). 
%Massive cosmological spectroscopic surveys provide a unique way to probe the 3D distribution of large-scale structures in the Universe, which can be used to measure the expansion history and growth of structures, thus constraining the Hubble parameter, as well as the properties of different cosmological components, such as the Non-Gaussianity and neutrino mass.
As the product of spectroscopic surveys, the database for the 3D positions of galaxies and quasars has been growing rapidly. In the past decade, surveys like the SDSS-III Baryon Oscillation Spectroscopic Survey \citep[BOSS;][]{boss} and SDSS-IV extended BOSS \citep[eBOSS;][]{eboss} have measured millions of spectra. The ongoing Dark Energy Survey Instrument \citep[DESI;][]{desi} is expected to take over 30 million spectra in 5 years. The next-generation experiments, such as the MegaMapper \citep{megamapper}, MUltiplexed Survey Telescope\footnote{\url{https://must.astro.tsinghua.edu.cn}} (MUST), WideField Spectroscopic Telescope \citep[WST;][]{wst}, and the MaunaKea Spectroscopic Explorer \citep[MSE;][]{Maunekea}, are expected to be equipped with a large number of fibres (10--20k) thanks to the development of robotic fibre positioners, and are foreseen to increase the number of observed galaxy/quasar by another order of magnitude.

These observations will allow us to test the standard $\LCDM$ model of cosmology, with parameters constrained at sub-per-cent-level precision. The standard $\LCDM$ model has been able to explain a large number of observations, from the CMB to low-redshift galaxies. However, tensions between measurements have recently become more and more significant, notably the Hubble constant $H_0$ \citep{Riess_hubble,Freedman_hubble} and the growth parameter $\sigma_8$ \citep{sigma8_2013,sigma8_largescale}. The origin of these tensions could come from bias in our measurements, unknown systematics or be the sign of new physics \citep{H0Sunny,S8Sunny,hubble_solutions,hubble_bias}.
%There are alternative models to $\LCDM$, with additional parameters to constrain with observations, such as 
There are extensions of the standard $\LCDM$ model, e.g., varying the dark energy equation of state \citep[][]{DE_eos,Tripathi_2017}, the primordial non-Gaussianity \citep[NG;][]{NG_epsilon_2000,NG_Dalal2008}, the non-zero total neutrino mass \citep{thesis_neutrino}.

The primordial non-Gaussianity is a test to inflation scenario \citep{inflation_obs}, and the LSS of galaxies/quasars is controlled by the NG amplitude parameter $\fnl$ through a bias scale dependence \citep[][]{NG_inflations_2003,desjaques_transfer_fnl}.
% $b=b_G+\Delta b(\fnl,k)$
Therefore measuring the large-scale clustering of galaxies provides an opportunity to study the early Universe physics.

Oscillation experiments have shown that at least two families of neutrinos have non-zero mass \citep{neutrino_oscillations}, but they can only constrain the relative mass difference between families. %$\vert m_i^2-m_j^2\vert$. 
Normal and inverted hierarchy theories give different predictions of the minimal sum of neutrino mass respectively: $M_\nu\sim0.06$ eV, or $M_\nu\sim0.1$ eV \citep{neutrino_hierarchy}. As massive neutrinos constitute a small fraction of the energy density of the Universe, a range of cosmological probes can provide indirect evidence of their mass properties \citep{neutrino_cosmo}. For example, a joint analysis of the Planck cosmic microwave background (CMB) measurements, and the Baryon Oscillation Spectroscopic Survey (BOSS) galaxy clustering data,  have already put an upper limit $M_\nu<0.16$ eV at 95 per cent confidence level \citep{neutrino_planck_bosss}, and down to  $M_\nu<0.11$ eV with data from eBOSS \citep{neutrino_eboss_mesure}. Thus, the Universe appears to be an ideal laboratory for the measurement of the neutrino hierarchy.
However, it has been recently objected that this measurement depends on the $\LCDM$ model (which has started to exhibit weakness), and cannot categorically exclude a scenario, though a careful study on the dependence of these constraints on the $\LCDM$ model may be necessary \citep{neutrino_modele}. 

The aim of this article is first to forecast the accuracy of high redshift spectroscopic surveys of the next decade, on four cosmological aspects: BAO scales, the RSD effect, NG amplitude parameters, and the sum of neutrino mass. We use Fisher information matrix $F_{ij}$ \citep[][]{first_fisher} for this purpose, assuming its inverse is a typical covariance matrix of our parameters (which is true in the Gaussian case, and gives an upper limit in the general one). We will use the linear theory to evaluate this information matrix. Thus our forecasting method is not particularly innovative compared to modern Markov chain Monte Carlo \citep[MCMC;][]{MCMC}  methods, and the consideration of non-linearity. Our goal is rather to compare the different surveys, and to apply this simple formalism in the context of an optimisation of the parameters of a survey. In principle, it is also possible to perform Fisher forecasts for surveys probing $z\sim 1$, with a much higher tracer density. Nonetheless, the main interest of these surveys is to extract more information from small scales, e.g., by exploring the power spectrum up to $k_{\rm max} \sim 0.5 h$ Mpc$^{-1}$ modes. The linear model of the power spectrum that we adopt in this study is not able to describe these scales accurately. Therefore, forecasts for high-density surveys that focus on small-scale clustering are left for future work.

With the increase of the number of fibre, and as demonstrated by our forecasts, future surveys will be able to constrain  cosmological parameters such as BAO and RSD at the sub-per-cent level. Besides, the measurement of parameters beyond the standard model like $\fnl$ and $M_\nu$, remains a challenge as the error bars are comparable to the parameter values.  
That is why, in a second step, we produce a quantitative optimisation pipeline of the observation strategy of spectroscopic surveys, for the study of the parameters $\fnl$ and $M_\nu$ as we find their constraints still need improvement despite their large number of fibres and large cosmic volume. It could be used for the design of future surveys, but it also provides a point of comparison between the expected accuracy of some surveys and the technically optimal one. To do so, we will present a rather general model, of a high redshift spectroscopic survey. 
We model in a simplified way the properties of observational targets, the specifications of the telescope, and the survey strategy.

This paper is organised as follows. In Section \ref{sec:survey}, we describe the future high redshift massive spectroscopic surveys in detail. The methodology used for the science forecasts is outlined in Section \ref{sec:method}, as well as our modelling for a general spectroscopic survey. We present the cosmological parameter constraints and the preferred improvements for future surveys in
Section \ref{sec:results}. Finally, we summarise our results in Section \ref{sec:conclusion}.

\section{Surveys}
\label{sec:survey}
In this section, we describe various spectroscopic galaxy surveys and cosmological probes considered in our forecasts. These surveys will observe emission line galaxies (ELGs) that are abundant up to redshift $\sim 2$ \citep{madau2014}, Lyman alpha emitter galaxies (LAEs),  Lyman break galaxies (LBGs), and BX galaxies \citep{BX} that can be observed up to redshift 5, or even higher in theory \citep{bias_luminosity}. 
%ELG are abundant up to redshift $\sim 2$ \citep{madau2014}, LBG and LAE can be observed up to redshift 5, and theoretically even 6 or 7. 
Forecasts presented in this work do not include constraints on cosmological parameters coming from cosmic shear, HI intensity mapping and future CMB observations that will be included in upcoming surveys \citep{snowmass} such as Euclid \citep{Euclid}, DESI \citep{desi}, Puma \citep{PUMA}, HETDEX \citep{hetdex}.

Survey properties considered in our study are listed in Table \ref{tab:survey_prop}, and the associated densities are presented in Table \ref{tab:RSDBAO} in Section~\ref{sec:results}. In general, the number density of a given tracer $X$ can be (ideally) modelled by \begin{equation}
    n(z)=\int^{m_\text{max}}E(m)\phi _X(m,z)dm,\label{eq:Em_phi}
\end{equation} 
where $m$ is the apparent magnitude of the tracer, $m_\text{max}$ is the maximum apparent magnitude of the survey, $E(m)$ is the efficiency of observation and $\phi_X$ is the tracer luminosity function. If the efficiency is independent of the magnitude, it reduces to 
\begin{equation}
    n(z)=\text{e}_\text{ff}\cdot n_X(m_{\text{max},z}),
    \label{eq:eff_n}
\end{equation}
where $\text{e}_\text{ff}$ is the constant efficiency of the tracer, and $n_X$ is the theoretical tracer density (cf Section \ref{sec:tracers}). We consider a redshift uncertainty $\sigma_z/(1+z)=0.001$ \citep{Maunekea} for every survey. Indeed the redshift resolution is assumed to be similar, but slightly more conservative than DESI one \citep{desi} to ensure that it is a reasonable estimation across our wavelength range.

\begin{table*}
  \begin{center}\label{tab:survey prop}
    \caption{Properties of future surveys, including their tracers, the redshift range and sky coverage, the number of fibres, the telescope diameter and the telescope location. Most of these properties are not definitive yet.}
    \label{tab:survey_prop}
    \begin{tabular}{ 
    |c|c|c|c|c|c|c|}
        \hline
         \multirow{2}{3em}{\textbf{Survey}}& \multirow{2}{3em}{\textbf{Tracer}} & \textbf{Redshift}& \textbf{Sky Coverage} & \textbf{Fibre} & \textbf{Telescope Size} &\multirow{2}{3em}{\textbf{Location}}\\
         &&\textbf{Range}&(deg$^2$)&\textbf{Number}&(m)&\\
        \hline
        \multirow{2}{5em}{MegaMapper}&\multirow{2}{2em}{LBG}&\multirow{2}{5em}{$2<z<5$}&\multirow{2}{3em}{14,000}&\multirow{2}{3em}{20,000}&\multirow{2}{1em}{6.5}&Las Campanas Observatory\\
        &&&&&&Chile\\
        \hline
        \multirow{2}{2em}{MSE}&ELG&$1.6<z<2.4$  &\multirow{2}{3em}{10,000}&\multirow{2}{2em}{4,332}&\multirow{2}{2em}{11.25}&Mauna Kea Observatories\\
        &LBG& $2.4<z<4$&&&&Hawaii, USA\\
        \hline
        WST &\multirow{2}{2em}{LBG}&\multirow{2}{5em}{$2<z<5$}&\multirow{2}{3em}{15,000}&20,000&\multirow{2}{2em}{11.4}& \multirow{2}{7em}{Northern Chile}\\
        (or Spectel)&&&&--60,000&&\\
        \hline
        \multirow{2}{3em}{MUST}&LBG +BX& \multirow{2}{5em}{$2<z<4$}
        & 9,000&10,000
        &\multirow{2}{1em}{6.5}& Lenghu, Qinghai province\\
        &--LBG+BX+LAE&&--15,000&--20,000 && China\\
        \hline
    \end{tabular}
  \end{center}
\end{table*}

\subsection{MegaMapper}
\label{sec:mega}
MegaMapper \citep{megamapper} is a spectroscopic instrument that will be located at the Las Campanas observatory in the southern hemisphere. It would target LBG at high redshift $2 < z < 5$, covering 14,000 square degrees of the sky. Its
6.5-meter telescope, 20,000 fibres, and a five-year observation period would yield a galaxy
number density $n> 10^{-4}h^
3 \text{Mpc}^{-3}$ across its redshift range. For the property of the fiducial LBG sample, we use the values in Table 2 of \citet{forecast_z_2_5}. These values are compatible with the model given by equation \eqref{eq:eff_n}, with $n_\text{LBG}$ an idealised density distribution introduced in subsection \ref{sec:tracers} (Eq.~\eqref{nLBG}),  $\text{e}_\text{ff}=0.4$ for $z<4$ and $\text{e}_\text{ff}=0.9$ for $4<z<5$ \citep[see Figure 4 of][]{white_same}, and $m_\text{max}=24.5$. 

\subsection{MSE}
\label{sec:MSE}
The MaunaKea Spectroscopic Explorer \citep[MSE;][]{Maunekea} will be located in Hawaii in the Northern hemisphere, 
probing over 10,000 square degrees. It will couple an 11.25-meter mirror with a 1.5-square-degree
field of view (FoV) to 4000 fibres, feeding to spectrographs that cover 360 to 1300 nm. This
design enables the detection of ELGs at $1.6< z <2.4$, and LBGs at $2.4 < z < 4$. The  exposure time is 1800s.%, the redshift precision $\sigma_z/(1+z)=0.001$ should be accomplished. 

The ELG number density shown in Table~\ref{tab:RSDBAO} is taken from  \citet{Maunekea}.
For LBG we will assume a model described by equation  \eqref{eq:Em_phi}. 
We estimate the efficiency thanks to Figure 2 of \citet{Maunekea}, assuming that $40$ per cent of LBG have Equivalent Width values EW$<0$, $30$ per cent have $0<$EW$<20$ and $30$ per cent have EW$>20$, and averaging over EW \footnote{We are averaging over the three templates $E(m)=0.4E(m\vert \text{EW}<0)+0.3E(m\vert 0<\text{EW}<20)+0.3E(m\vert 20<\text{EW})$}. The effective efficiency law is then given by $E(m)=-0.18m+4.8$. 
Given the approximate efficiency rate of 0.5, and the required observed density ($n=10^{-4}h^3/\rm{Mpc}^3$), 1400 fibres/deg$^2$ will be allocated to LBG observations, restricting to a maximum magnitude $m_\text{max}=24.2$ \citep{Maunekea}. We introduced our LBG luminosity function model in subsection \ref{sec:tracers}.

\subsection{MUST}
The MUltiplex Survey Telescope\footnote{\url{https://must.astro.tsinghua.edu.cn}} (MUST) is a future 6.5-meter telescope (with a 7  square degree FoV) located in China, in the Northern hemisphere. Its target can be either LBG+BX (LBGX), or a combination of LBGX+LAE. Since the exact survey design is still in flux, our forecast supposes its redshift range to be $2<z<4$, with a sky coverage between 9,000 and 15,000 deg$^2$, and fibre numbers to be either  10,000 or 20,000.  

\subsection{A WST-like NTL survey}
The WideField Spectroscopic Telescope\footnote{previously named Spectel, \url{https://www.wstelescope.com/}} \citep[WST;][]{wst} is a proposed spectroscopic survey in the southern hemisphere that would couple an 11.4m dish (with a 5 square degree FoV) and 20,000--60,000 fibres, enabling more than a hundred million of fibre exposures (each
$\geq$ 4,000 seconds long) over its survey period. Its design would permit observations of LBGs and LAEs up to redshift 5, with number densities 2 to 5 times of those for a MegaMapper-like survey. Since the exact design of this survey is a work in progress, we also explore several possible survey parameters. 

For the forecast, we consider a similar survey to MegaMapper, with a sky coverage of 15,000 deg$^2$ for LBG at redshift 2 to 5 and each tracer can be observed for a period as long as needed until it reaches the required spectrum quality. We model it with an efficiency $e_{\text{ff}}=0.9$ relative to the theoretical tracer density (cf. equation \eqref{eq:eff_n}), but with different maximum magnitudes -- 24.2, 24.5, and 25 -- depending on the final fibre number. This might correspond to 20,000--40,000--100,000 fibres for 5 years of observation\footnote{Of course we do not expect  surveys to have 100,000 fibres in the next decay. This forecast rather serves as an upper limit.}. %: $n(z)=0.9n_\text{LBG}(z,m_\text{max})$.
%We take $\sigma_z=0.001(1+z)$. 
This forecast somehow represents a cosmological limit on the achievable parameters accuracy, since we are assuming an efficiency very close to 1. As we do not really take into account the final properties of WST in our modelling, we will refer to this fiducial survey as a NTL survey (a No-Time-Limit survey) in Section~\ref{sec:results}.

\subsection{A General Survey}
\label{sec:GSS}
We consider also a high redshift ($z>2$) general spectroscopic survey with a modelling of the survey settings, in order to explore the optimal strategy that yields the tightest constraints of chosen cosmological parameters. 
In a first step, we assume an LBG survey lasting 5 years, based on a 10-meter telescope equipped with 20,000 fibres.
Since LBGs are abundant mostly at $z\gtrsim 2$ \citep{bias_luminosity}, we consider a redshift window $[z_\text{min},z_\text{max}$] that always starts with $z_\text{min}=2$.
The survey volume will be thus described by the fraction of the survey sky coverage $f_\text{sky}$\footnote{$f_\text{sky}=1$ corresponds to the full sky, 41,253 square degrees.} and the redshift span $\Delta z=z_\text{max}-z_\text{min}$.
%We will also vary this 5 year observation time, and explain how to extrapolate our results to those of a telescope with a different aperture, with a different survey duration. 
In a second step, we will also vary the `observation capacity' defined as the product of the survey duration and the fibre number, in to extend this model to a larger variety of spectroscopic telescopes and to highlight the improvement of the measurements with the available technology. 
We detailed the modelling of such a survey in Section~\ref{sec:GS}.

\section{Methodology}
\label{sec:method}
In this section, we first describe some properties of observed galaxies. Then we introduce the commonly used Fisher matrix forecasting technique. In Section~\ref{sec:cosmo_param} we specify the BAO, RSD, non-Gaussianity and Neutrino mass forecast strategies.  We then  introduce our modelling of a general survey in \ref{sec:GS}. 

\subsection{Observed power spectrum}
The power spectrum of a dark matter tracer $X$ is related to the theoretical matter power spectrum $P_\text{m}(k,z)$ with
\begin{equation}
    P_X(k,\mu,z)=(b_X(z)+f(z)\mu^2)^2P_\text{m}(k,z),
\end{equation}
with $b_X$ being the tracer bias, $f$ being the growth rate, and $\mu$ being the cosine between the line of sight and the 3d mode $\boldsymbol{k}$. To take into account the error in the redshift measurement $\sigma_z/(1+z)$ that propagates to an error in the radial distance via $\sigma_\chi=\sigma_z c/H(z)$, we multiply the power spectrum by a factor $\exp(-k^2\mu^2\sigma_\chi^2)$ \citep{white_same}.

We also introduce the cross-power spectrum of two different tracers $A$ and $B$ following \citet{power_spectrum_cross} as 
\begin{equation}
    P_{AB}(k,\mu,z)=(b_A(z)+f(z)\mu^2)(b_B(z)+f(z)\mu^2)P_\text{m}(k,z),
\end{equation}
where $b_A$ and $b_B$ are biases of tracer A and B respectively. 
We report in Table~\ref{tab:prior_fiducial} fiducial values of six standard cosmological parameters used in this work, along with the extension model parameter $M_\nu$.

\begin{table}
\caption{Fiducial values of cosmological parameters and their Planck Gaussian prior. NG amplitude $\fnl$ is neglected except for the NG forecast.}
\begin{tabular}{ccccccc}
\hline
$h$ & $\omega_b$& $\omega_c$&$n_\text{s}$&$\tau$&$\ln(A_\text{s})$&$M_\nu$ (eV)\\
\hline
\multicolumn{7}{c}{\textbf{Fiducial values}} \vspace{0.1cm}\\

0.677&0.02247&0.1192&0.9675&0.056&-13.073&0.06 \\
\hline
\multicolumn{7}{c}{\textbf{Planck half-width Gaussian prior}}\vspace{0.1cm} \\
0.0054&$0.00015$&$0.0012$&$0.0042$&$0.007$&$0.015$ &0.5\\
\hline
\end{tabular}
\label{tab:prior_fiducial}
\end{table}

Power spectrum will be evaluated using \textsc{CAMB}\footnote{\url{https://camb.readthedocs.io/en/latest/}} \citep{camb} and \textsc{pyccl}\footnote{\url{https://ccl.readthedocs.io/en/latest/}} \citep{Chisari_2019}.

\subsection{Bias and luminosity function}
\label{sec:tracers}

For ELG, we assume a constant clustering amplitude, based on the analysis of DESI-selected samples in the DEEP2 data \citep{desi}. In that case, the bias can be approximated as $b_\text{ELG}=0.8\times D(0)/D(z)$ with $D$ the growth function. 
Factor 0.8 is chosen to be slightly lower than that of DESI \citep[0.84; see][]{desi} and eBOSS \citep[1; see][]{eboss}, as we consider fainter ELGs in this study.  
 
We model LBG and LAE bias following the parametrization of \citet{bias_luminosity} using
\begin{equation}
b_\text{LBG/LAE}(z,m)=A(m)(1+z)+B(m)(1+z)^2,
\end{equation}
with $A(m)=-0.98(m-25)+0.11$ and $B(m)=0.12(m-25)+0.17$, $m$ being the apparent magnitude. 
We assume that fainter galaxies contribute more, since the galaxy abundance grows as the magnitude increases, and reduce the bias to a one-parameter function $b(z,m)\approx b(z,m_\text{max})$.
For samples with a large magnitude band, we might separate it into subsamples of different maximal magnitudes, and adopt a multi-tracer approach. 

For the aim of our study, we need to evaluate the LBG density function.
An idealized number density is modelled by 
\begin{equation}
    n_\text{LBG}=\int_{-\infty}^{M_c}\phi(M)dM,
\end{equation}
with $\phi$ the luminosity function \citep{ bias_luminosity,white_same}, and $M$ the absolute magnitude. We use the Schechter model \citep{1976ApJ...203..297S} for the luminosity function:
\begin{align}
    \phi(M)=\frac{\ln 10}{2.5}  \phi^\star 10^{-0.4(1+\alpha)(M-M^\star)}\exp \left( -10^{-0.4(M-M^\star)}\right)
\end{align}
with $\alpha$, $M^\star$ and $\phi^\star$ listed in Table 3 of \citet{bias_luminosity}. The absolute magnitude cutoff $M_c$ of galaxies at redshift $z$ with apparent magnitude being $m_\text{max}$ is determined as 
\begin{equation}
    M_c(m_\text{max})=m_\text{max}-5\log_{10}\left(\frac{D_\text{L}(z)}{10\text{pc}}\right)+2.5\log_{10}(1+z),
\end{equation}
with $D_\text{L}(z)$ the luminosity distance.
As the observation depends rather on the apparent magnitude $m$ than the absolute magnitude $M$, we rewrite, the density equation, with $\phi(m,z)=\phi(M(m,z))$, using $dM/dm=1$, as 
\begin{equation}
\label{nLBG}
    n_\text{LBG}(z)=\int^{m_\text{max}} \phi(m,z)dm.
\end{equation}
In the rest of the study, we will use this last equation formalism and refer to $m$ as `magnitude' hereafter.

\subsection{Fisher Matrix}
For a set of cosmological parameters $\{p_i\}$, the diagonal coefficient of the inverse of its Fisher matrix $\mathcal{F}_{ij}$ gives an upper bound on the variance of each parameter: $\sigma_i^2\ge (\mathcal{F})^{-1}_{ii}$ according to the Cramer-Rao inequality (with equality for Gaussian likelihood). We follow the same steps as \citet{eBOSSforecast} and considered the Fisher matrix 
\begin{align}
\mathcal{F}_{ij}&=\frac{V_{\text{sur}}}{4\pi^2}\int_{-1}^{+1}d\mu\int_{k_\text{min}}^{k_\text{max}}k^2dk F_{ij}(k,\mu),\\
F_{ij}&=\frac{1}{2}\text{Tr}(\p_{p_i} \mathbf{C}\mathbf{C}^{-1}\p_{p_j}\mathbf{C}\mathbf{C}^{-1}),
\end{align}
with $V_\text{sur}$ the comoving volume of the survey, and $\mathbf{C}$ the data covariance matrix. The integration bound $k_\text{min}$ depends on the survey volume and corresponds to the maximal length, while $k_\text{max}$ depends on the accuracy of the theoretical model on non-linear scales and on the shot noise. We take by default
\begin{equation}
k_\text{min}=\frac{2\pi}{V_{\text{sur}}^{1/3}} [\text{$h\,$Mpc}^{-1}],~k_\text{max}=\frac{0.1D(0)}{D(z)}[\text{$h\,$Mpc}^{-1}].\label{eq:kmax}
\end{equation}
Since the neutrino mass, and more generally many new physics properties, such as the nature of gravity, are significantly encoded at small scales, we will also consider an `optimistic' integration bound with $k_\text{max}=0.3h\rm{Mpc}^{-1}$. This is motivated by both the reduction of shot noise in future data survey, and expected progress in theoretical understanding and modelling of non-linearities.

\subsubsection{One Tracer}

When targets are the same type of tracer (MegaMapper LBGs for example), $C$ is a $1\times 1$ matrix. 
We take into account the tracer distribution discreetness by adding a Poissonian shot noise that scales as the inverse of the number density $1/n$ as 
\begin{equation}
    C=P+1/n,
\end{equation}
where $P$ is the power spectrum of this tracer. 
Thus, the Fisher matrix is
\begin{equation}
    F_{ij}=\frac{1}{2}\left(\frac{nP}{nP+1}\right)^2\frac{\p \ln P}{\p p_i}\frac{\p \ln P}{\p p_j}.
\end{equation}
We will report the parameter $nP(k=0.14,\mu=0.6)$ for different surveys, an approximate centre-of-weight point for BAO measurements.to give a qualitative description of the noise level, following \cite{desi}.
\subsubsection{Two tracers}
\label{sec:fisher_2tracers}
For two tracers (LBG+LAE in MUST for example), $C$ is now a $2\times2$ matrix, and under the same assumption,
\begin{equation}
    \mathbf{C}=\begin{bmatrix} 
	P_{AA}+\frac{1}{n_A} & P_{AB}\\
	P_{AB} &  P_{BB}+\frac{1}{n_B} \\
	\end{bmatrix}.
\end{equation}
where $P_{AA}$ and $P_{BB}$ are auto-power spectra of tracer A and B, $P_{AB}$ is the cross-power spectrum of tracer A and B. An explicit expression for the Fisher matrix in the 2 tracers case is given in Appendix A of \citet{eBOSSforecast} as combinations of $\frac{\p \ln P_T}{\p p_i}$ with $T=A,B,AB$. In the case of two independent tracers, $\mathbf{C}$ is diagonal and the two-tracers Fisher matrix is equal to the sum of the two one-tracer matrix.  
Similarly, as in the one-tracer case, we will report the parameter $\sum n_iP(k=0.14,\mu=0.6)$.

For surveys with extremely massive data sets, whose galaxy apparent magnitudes are spread over a wide band, we separate them into sub-samples with different magnitude ranges (typically $\{-\infty;~24.5\}$, $\{24.5;~24.8\}$). Then we adopt a two-tracer forecast approach. 
Such a process is mainly motivated by the non-Gaussianity forecast, since $\delta b\propto f_\text{NL}b_G$, cf. section \ref{sec:fnl_theory}. Indeed the bias decreased with the magnitude, and a one tracer approach with $b=b(m_\text{max})$ assumption  artificially reduces the sensitivity to NG.   

\subsubsection{Complementary data sets}
\label{subsec:complementary data}
To combine constraints from two independent surveys\footnote{Surveys with non-overlapping redshift ranges, sky coverages or independent tracers.} A and B that aims to measure the same set of cosmological parameters, one simply adds their Fisher matrix $F_{ij}=F^A_{ij}+F^B_{ij}$. Furthermore, to introduce priors from complementary data sets (such as Planck CMB) that have measured parameters $\{p_i\}$ with accuracies $\{\sigma_i\}$, one simply adds to the Fisher matrix $P_{ij}=\delta_{ij}/\sigma_i^{2}$ (assuming Gaussian uncertainties).

\subsubsection{Redshift bins}
\label{sec:redshift_binnung}
Forecasts for a survey that covers a large redshift range $[z_\text{min},z_\text{max}]$, have to take into account the redshift dependence of parameters. There are three approaches to do so:
\begin{itemize}[wide, labelwidth=!,itemindent=!,labelindent=0pt, leftmargin=0em, itemsep=.1cm, parsep=0pt]
    \item Split the survey volume into redshift bins $\{z_k\}$ with separation $dz_k$, and present the forecast for each bin separately,  
    \item Separate the survey volume into redshift bins $\{z_k\}$ with separation $dz_k$, and sum the fisher matrices, neglecting cross correlation between bins: $F_{ij}=\sum_k F^{z_k}_{ij}$,
    \item Consider one redshift bin, with an effective redshift $z_\text{eff}$  ( so with this approach one neglects redshift dependence of the parameters, bias and density). 
\end{itemize}

Following \citet{white_same}, the effective redshift of of a subsample is calculated with 
\begin{equation}
    z_\text{eff}=\frac{\int dz H^2(z)\chi^2(z)(d\chi/dz)^3n^2(z)z}{\int dz H^2(z)\chi^2(z)(d\chi/dz)^3n^2(z)}.
\end{equation}
None of these approaches is flawless, and we will choose the best one for different purposes in the following study. \citet{cross_correlatin_bins} has implemented a multi-bins approach, and show that in some case the cross-correlation between bins modifies the forecast up to 10-20 per cent. Nonetheless, \citet{white_same} has shown that for these high-redshift galaxy surveys, the correction was negligible ($<10$ per cent, cf. Appendix B of their paper).
\citet{optimal_redshift_weight} have calculated the optimal redshift weighting scheme for the BOSS survey and a similar algorithm can be implemented for future surveys, but this is beyond the scope of this study.

\subsection{Cosmological Parameters}
\label{sec:cosmo_param}
\subsubsection{BAO}
For the BAO forecast, the two parameters are $\ln(D_A/s)$ and $\ln(sH)$, with $s $ the sound horizon, $D_A$ the angular distance, and $H$ the Hubble parameter. We assume to have a very good measurement of $s$ from CMB, so that $\sigma(D_A/s)=\sigma(D_A)/s$ and $\sigma(sH)=s\sigma(H)$, thus:
\begin{equation}
    \sigma(\ln(sH))=\frac{\sigma(H)}{H};~~ \sigma(\ln(D_A/s))=\frac{\sigma(D_A)}{D_A}.
\end{equation}
The distance error on both of the parameters is derived with the \citet{BAO_simple} approximation of the Fisher matrix,
\begin{equation}
\begin{aligned}
    \mathcal{F}_{ij}=&V_{\text{sur}}A_0^2\int_0^1d\mu \int_0^\infty  dk \Bigg\{f_i(\mu )f_j(\mu ) k^2 \\
    \times &\frac{\exp\left( -2\left(k\Sigma_s\right)^{1.4}\right)}{\left(\frac{P(k)}{P(0.2)}+\frac{1}{nP(0.2)}\right)^2}\exp\left(-k^2(1-\mu^2)\Sigma_\perp^2-k^2\mu^2\Sigma_\parallel ^2\right)\Bigg\},
\end{aligned}
\end{equation}
where
\begin{equation}
     f_i(\mu) = \left\{ \begin{array}{ll}
         \mu^2-1 & \mbox{if $i=1$};\\
        \mu^2 & \mbox{if $i=2$}.\end{array} \right. 
\end{equation}

$\Sigma_\parallel$ and $\Sigma_\perp$ are the root mean square displacement along and perpendicular to the line of sight. $\Sigma_\parallel =\Sigma_0D(z)(1+f(z))$ 
 and $\Sigma_\perp=\Sigma_0 D(z)$ with $\Sigma_0=10.4\sigma_8\,h^{-1}\rm Mpc$. \footnote{ The value in \citet{BAO_simple} for $\Sigma_0$ is different since we chose to work with $D(z)$ instead of $G(z)$,and we chose a different $\sigma_8$ value.} The Silk-damping effect is included with the Silk-damping scale $\Sigma_s$, expressed in $h^{-1}\rm{Mpc}$ via
\begin{equation}
     \Sigma^{-1}=1.6\left(\Omega_bh^2\right)^{0.52}\left(\Omega_m h^2\right)^{0.73}\left[1+\left(10.4\Omega_mh^2\right)^{-0.95}\right]h^{-1}.
\end{equation}
We assume a reduction of the BAO damping scale by a factor 0.5 w.r.t. the value from \citet{BAO_simple}, following section 4.1 of \citet{Andreu_neutrino}. We fixed $A_0=0.55$, the WMAP3 value given in \citet{BAO_simple}.\footnote{ We have tried varying $A_0$ between 0.45 and 0.6, the resulting difference on $\sigma(H)$ and $\sigma(D_A)$ w.r.t.  $A_0=0.55$ was about $10$ per cent, which is not significant for our work.  }

Here for BAO measurements in the two-tracer case, we will assume two independent measurements, and simply sum the two fisher matrices. %The noise parameter is $\sum n_iP_i(k=0.14,\mu=0.6)$.

\subsubsection{RSD effects}
We follow \citet{rsd} for the Redshift Space Distortions forecast. 
For one tracer, we rewrite the equation of the power spectrum:
\begin{equation}
    P(k,z)=\left(b(z)\sigma_8(z)+f(z)\sigma_8(z)\mu^2\right)^2\frac{P_\text{m}(k,z=0)}{\sigma_8(z=0)^2}.
\end{equation}
We introduce our parameters: $\ln[b(z_i)\sigma_8(z_i)]$ and $\ln[f(z_i)\sigma_8(z_i)]$.
For simplicity, we drop the explicit redshift dependence. In this case, the derivative of power spectrum w.r.t. parameters are
\begin{align}
    \frac{\p \ln P}{\p \ln(b\sigma_8)}&=\frac{2b\sigma_8}{b\sigma_8+f\sigma_8\mu^2}\\
    \frac{\p \ln P}{\p \ln(f\sigma_8)}&=\frac{2f\sigma_8\mu^2}{b\sigma_8+f\sigma_8\mu^2}.
\end{align}

Everything is similar for the case of two tracers A and B, except that we have 3 sets of parameters: $\ln[b_A(z_i)\sigma_8(z_i)]$, $\ln[b_B(z_i)\sigma_8(z_i)]$ and $\ln[f(z_i)\sigma_8(z_i)]$, with additional derivatives:
\begin{align}
    &\frac{\p \ln P_A}{\p \ln(b_B\sigma_8)}=\frac{\p \ln P_B}{\p \ln(b_A\sigma_8)}=0,\\
    &\frac{\p \ln P_{AB}}{\p \ln(b_x\sigma_8)}=\frac{b_x\sigma_8}{b_x\sigma_8+f\sigma_8\mu^2}\\
    &\frac{\p \ln P_{AB}}{\p \ln(f\sigma_8)}=f\sigma_8\mu^2\left(\frac{1}{b_a\sigma_8+f\sigma_8\mu^2}+\frac{1}{b_b\sigma_8+f\sigma_8\mu^2}\right),
\end{align}
where $x=A,B$. 
\label{sec:RSD}

\subsubsection{Non Gaussianity}
\label{sec:fnl_theory}
In most NG model \citep{NG_epsilon_2000,NG_inflations_2003,NG_Dalal2008,desjaques_transfer_fnl}, the Bardeen potential is assumed to contain a quadratic Gaussian field $\phi$ contribution $\Phi=\phi+\fnl(\phi^2-\langle\phi\rangle^2)$. The quadratic term induces a non-Gaussian perturbation to the bias: $b(k,z)=b_G(z)+\Delta b(k,z)$ with:
\begin{equation}
    \Delta b=3\fnl(b_G-p) \delta_c \frac{\Omega_m}{k^2T(k)D(z)}\left(\frac{H_0}{c}\right)^2.
    \label{eq:pparam}
\end{equation}
$\fnl$ is the NG coupling to evaluate, $T(k)$ is the transfer function (with $k^2T(k)$ normalized to 1 at large scales), and $p$ is a number that theoretically depends on the tracer type, and was introduced to show deviations from the original model of \citet{NG_Dalal2008}. We take $\fnl=0$ as a fiducial value. Since $p$  is not well characterised yet, we will assume $p=1$ for all the different tracers. The lack of knowledge on $p$ value will be discussed in Section~\ref{sec:results}. We consider two parameters for the forecast $\{b_g,\fnl\}$, with
\begin{equation}
    \frac{\p P}{\p \fnl}=2\frac{\partial \Delta b}{\partial \fnl}(b_G+\Delta b +f \mu^2)P_\text{m}(k).
\end{equation}
For the generalization to two tracers, there is an additional derivative
\begin{equation}
    \frac{\p P_{AB}}{\p f_{\text{NL}}}=\frac{\p \Delta b_a}{\p f_{\text{NL}}}(b_b+\Delta b_b +f\mu^2)P_\text{m}(k)+\{b\leftrightarrow a\}
\end{equation}

\citet{NG_bispectrim,forecast_z_2_5} have suggested that by using bispectrum in addition to the power spectrum, one might be able to  reach $\sigma(\fnl)\sim 0.1$. The modelling of bispectrum observations is relatively complex, and this high-precision constraint requires indeed more theoretical development from the modelling side, as well as better understanding of systematics effects \citep{bispectrum_difficult}.  
%indeed relatively unrealistic in practice.
Our study is far too simplistic to address these issues, so we leave it for future studies.

\subsubsection{Neutrino mass}
\label{sec:neutrino_limit}
The evaluation of the sum of the neutrino mass is an active subject, both in cosmology and particle physics. We adopt a linear-power-spectrum Fisher approach. It may seem too simple to the current standard algorithm, which consists of forecasting with an MCMC and non-linear corrections in the model \citep[e.g., ][]{white_same}. However, our aim in this paper is to study the change of constraints w.r.t. that of observational parameters in spectroscopic surveys. Furthermore, even the most advanced approaches do not agree with each other \citep[an issue discussed in][]{thesis_neutrino}.%, and in the past, constraints from observations were not as good as their forecasts \citep{Andreu_neutrino,eBOSSforecast,neutrino_eboss_mesure}. 
Indeed, we expect that the difference between our power spectra and those provided by a more complex $M_\nu$ will be relatively similar for all surveys and our conclusion should not be affected.  

%\subsubsection{Formalism}
\label{sec:NMform}
The sum of neutrino mass is denoted as
\begin{equation}
    M_\nu=\sum_\nu m_\nu.
\end{equation}
We consider a cosmology described by the six standards cosmological parameters and $M_\nu$,
\begin{equation}
    \{p_i\}=\{M_\nu,H_0,\omega_c,\omega_b,\tau,n_s,A_s\}.
\end{equation}
To evaluate the derivative of the power spectrum w.r.t. these parameters, we use a four-point estimate (we drop the $k,z,\mu$ dependency here):
\begin{equation}
    \frac{\p P}{\p \theta}\vert_{\theta_\text{fid}}\sim\frac{-P(\theta+2\delta \theta)+8P(\theta+\delta \theta)-8P(\theta-\delta \theta)+P(\theta-2\delta \theta)}{12\delta\theta}
\end{equation}
We use $\delta \theta/\theta=0.01$ except for $\delta \tau/\tau=0.5$ and $\delta M_\nu/M_\nu=0.05$ for numerical reasons. Indeed, we want the power spectrum variation (for every step) to be much larger than this numerical noise induced by solving Boltzmann equations. Thus for parameters which only have a small impact on the matter power spectrum (the neutrino mass $M_\nu$, and the optical depth $\tau$), we take a larger parameter step size. We pay particular attention to the numerical stability and convergence of the derivatives w.r.t. those steps. We fix the neutrino hierarchy as it is degenerated in CAMB, for numerical error purposes.  
 
One subtlety not always mentioned is that the power spectrum used in $M_\nu$ study only includes baryons and cold dark matter, and its associated growth rate $f$. Indeed neutrino perturbations do not contribute to the formation of galaxies and haloes \citep{thesis_neutrino}. We focus on the neutrino mass parameter and marginalised our Fisher matrix over all the other parameters.

\subsubsection{Additional data-sets for NM}
\label{sec:planck_desi}
We will add a prior, using Planck CMB constraints on our standard set of parameters: $(F^\text{Planck})_{ij}=\delta_{ij}/\sigma_i^2$ (Section~\ref{subsec:complementary data} and Table~\ref{tab:prior_fiducial}).
We will also consider the possibility of combining our forecast with DESI which has 3 tracers: ELG LRG and QSO. We split DESI ELG and QSO into several redshift bins and sum the corresponding Fisher matrix as
\begin{equation}
F^\text{DESI}=F^\text{DESI}_\text{LRG}+F^\text{DESI}_\text{ELG}+F^\text{DESI}_\text{QSO}, 
\end{equation}
neglecting the correlation between bins. We also neglect cross-correlation between tracers. Table~\ref{tab:desi_tracers} summarises the redshift binning of DESI tracers. Since our goal is not to derive the optimal bound, but rather to make comparisons among different surveys, we do not include additional information for BOSS or Euclid for example. The argument is similar to the one for neutrino mass modelling. 
%$F^\text{DESI}=\sum_{i\in T}\sum_jF^\text{i}_{z^i_j}$ 

\begin{table}
  \begin{center}
    \caption{The redshift slices and the mean number density at that redshift range for DESI tracers.}
    \label{tab:desi_tracers}
    \begin{tabular}{ |c|c|c|}
        \hline
        \multirow{2}{3em}{\textbf{Tracers}}&\textbf{Redshift}&$n$\\
        &\textbf{Range}&($10^{-4}\,h^3\,\rm{Mpc}^{-3}$)\\
        \hline
        LRG&0.65-1.05&3.0\\
        \hline
        \multirow{3}{2em}{ELG}&0.75-1.05&9.8\\
        &1.05-1.35&4.5\\
        &1.35-1.65&1.3\\
        \hline
        \multirow{2}{2em}{QSO}&1.96-2.43&0.17\\
        &2.43-3.55&0.063\\
        \hline
    \end{tabular}
  \end{center}
\end{table}

\subsection{High-Redshift Survey Modelling}
\label{sec:GS}
In this section we model a general survey (cf. section \ref{sec:GSS}), keeping engineering and observational parameters as variables. We will then investigate the survey characteristics to get the best constraint on parameters $f_\text{NL}$ and $M_\nu$.

\subsubsection{Time to observe a single galaxy}
The Signal-to-Noise Ratio (SNR) describes how well a source is measured by an instrument and is given by the CCD equation,
\begin{equation}
    S/N \propto \frac{I(m) S_\text{tel} t}{\sqrt{\epsilon_\text{source}+\epsilon_\text{sky}+\epsilon_{\text{read}}}},
\end{equation}
with $S_\text{tel}$ the telescope surface, $t$ the time of observation and $I(m)$ the luminous flux from the source. $\epsilon_\text{source}$, $\epsilon_\text{sky}$ and $\epsilon_{\text{read}}$ are the source noise, sky noise and read-out noise. LBG being faint galaxies, $\epsilon_\text{sky}$ will be the dominant term. It scales as $\epsilon_\text{sky}\propto S_{\rm tel}\, t$. Thus we have
\begin{equation}
S/N\propto I(m) \sqrt{ S_{\rm tel} t}    .
\label{eq:CCD}
\end{equation}

During a telescope operation phase, the time for observing an object is fixed to the exposure time $t_\text{exp}$, independently of its apparent magnitude. After a first exposure, the spectrum might still be too noisy to identify lines for determining the redshift.   
That defines the first-exposure magnitude efficiency $p(m\vert t_\text{exp})$ which is the probability of getting a redshift-identifiable spectrum during $t_\text{exp}$. 

\subsubsection{One Exposure}
\label{sec:one_exp}
For a single exposure of $t_\text{exp}=1800\,s$, with a telescope of diameter $\sim10\,m$, we assume that
\begin{equation}
  p(m\vert t_\text{exp})=E(m),
\end{equation}
where $E(m)$ is the observational efficiency introduced in section~\ref{sec:MSE} for the modelling of MSE efficiency \citep{Maunekea}.
This law depends on the exposure time $t_\text{exp}$, the telescope surface $S_{\text{tel}}$, the maximum tolerable redshift error $\vert z_\text{meas}-z_\text{real}\vert=0.001(1+z_\text{real})$, and spectra simulated with the MSE exposure time calculator given the redshift finder PandoraEZ \citep{PandoraEZ}. Furthermore, this law is based on the assumptions of LBG properties that half of LBG have a detectable Ly$\alpha$ emission line, and the redshift of the other half can be estimated with their Ly$\alpha$ and Ly$\beta$ absorption features. 
Thus, except for the exposure time, and telescope surface, this law is not specific to MSE survey, and should in principle apply to other spectroscopic surveys.

\subsubsection{Multiple Exposures}
After a first exposure, if the spectrum is still too noisy, a target may get another exposure $t_\text{exp}$.

We assume that if one observes an object with an additional exposure after a first failure, the probability of measuring its redshift accurately from the stacked spectrum is 
\begin{equation}
p(m\vert 2t_\text{exp})=\sqrt{2} p(m\vert t_\text{exp}).    
\end{equation}
Indeed, in Appendix~\ref{sec:AppendixJiaxi}, we show that the efficiency of the ELG detection increases linearly with the SNR for eBOSS ELGs, up to a very good precision. Thus, we assume a similar trend for high redshift LBGs, i.e., the efficiency is proportional to the SNR, and scales with $\sqrt{t}$ according to Eq~\eqref{eq:CCD}. If we go even further and decided to attribute the third exposure of 2 failures, we assume $p(m\vert 3t_\text{exp})=\sqrt{3} p(m\vert t_\text{exp})$ and so on.

\subsubsection{Correcting Efficiency Bias}
\begin{figure}
    \centering
    \includegraphics[scale=0.55]{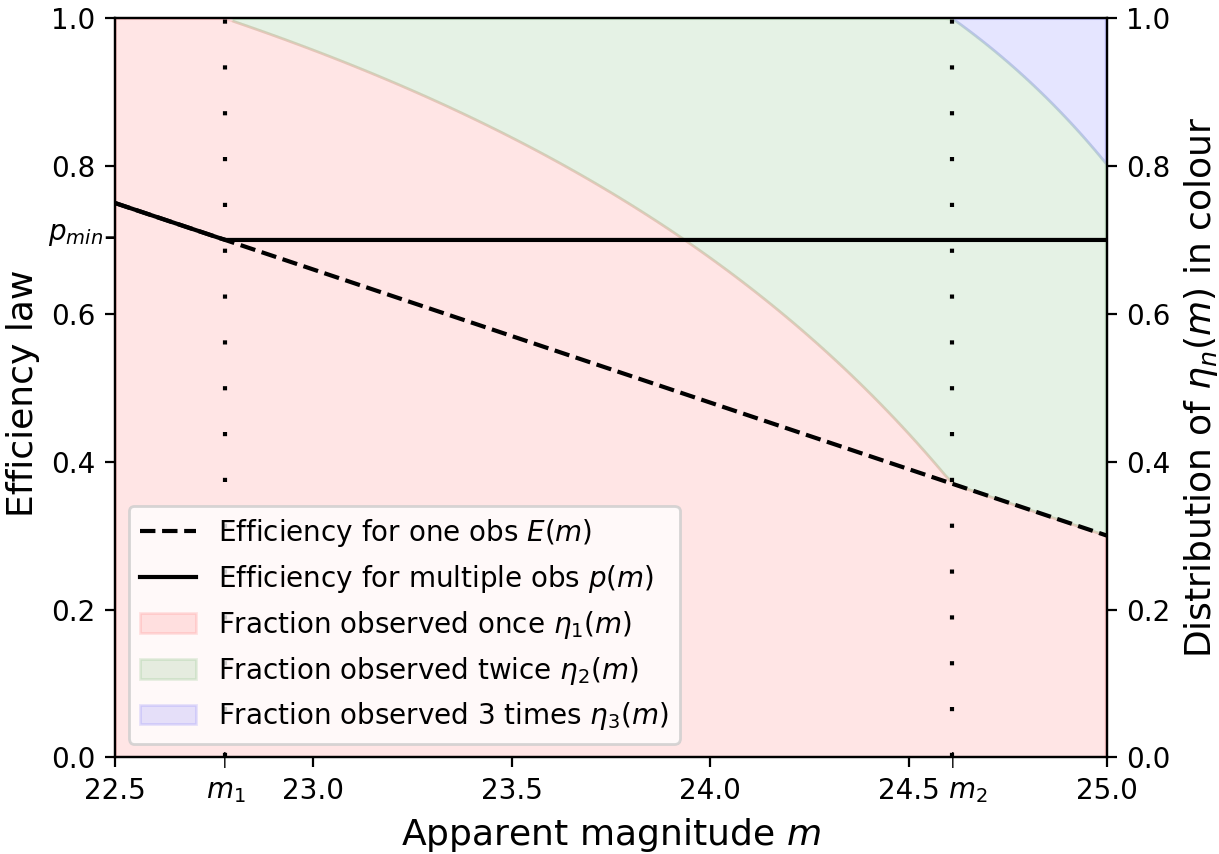}
    \caption{Efficiency law for a single exposure $E(m)$ (dashed line) and for multiple exposures $p(m)$ (solid line), with a minimal efficiency $p_\text{min}=0.7$ (reported on the y-axis on the left) as a function of apparent magnitude $m$. Colours represent the fraction of galaxies of magnitude $m$ that have been observed  $n$ times: $\eta_n(m)$, with $\sum_n \eta_n(m)=1$. Values of these fractions are deduced from the y-axis on the right. For e.g. at $m=25$, 30 per cent of galaxies are observed once, 50 per cent are observed twice, and 20 per cent are observed three times. We have also reported two particular magnitudes $m_1$ and $m_2$ by vertically pointed lines.}
    \label{fig:new_prob}
\end{figure}

The one-exposure efficiency of fainter tracers is significantly smaller than that of bright objects, and  one will underestimate the proportion of high-magnitude tracers. As a consequence, the sample will be biased to large $m_\text{max}$. 
That is why we define a minimal observational efficiency $p_\text{min}$ so that $\forall m<m_\text{max},~p(m)\geq p_\text{min} $. In practice  we decide to attribute a second exposure after the first failure to a certain fraction of object: $\eta_2$, with $0\leq \eta_2\leq 1-E(m)$ (we only assign a second observation if the first exposure failed). We also define the fraction of object observed once $\eta_1$, with $E(m)\leq \eta_1\leq 1$. Similarly, if observing all objects that failed during the first exposure twice does not compensate for the efficiency decreased, a fraction of galaxies $\eta_3$ might need a third observation. This procedure defines an efficiency law\footnote{We neglect the bias induced by the high efficiency of low-magnitude sources. }: $p(m)=\max (E(m),p_\text{min})$. As an example, in Figure \ref{fig:new_prob} we fix $p_{\text{min}}=0.7$ and we plot $p(m)$ (solid line), as well as $E(m)$ (dashed line), as functions of apparent magnitude $m$. For $m<m_1=22.8$, we have $p(m)=E(m)$;  $p(m)=p_{\text{min}}$ for higher magnitude. 
In the same figure, we also represent in colour the different fractions of galaxy observed $n$ times $\eta_n$, as a function of apparent magnitude $m$, with $\sum_n \eta_n(m)=1$. The fraction of objects that have been correctly observed during the first exposure\footnote{which is not $\eta_1(m)$} is equal to $E(m)$ by definition. For $m<m_1$, $E(m)>p_\text{min}$, and all the object are observed once: $\eta_1(m<m_1)=1$. For $m_1<m<m_2$, to compensate that $E(m)<p_\text{min}$, a fraction $\eta_2$ of galaxy is observed twice (the green shade). Then at $m=m_2$, we have $\eta_1=E(m_2)(=0.37)$, which means that all the objects whose observation failed during the first exposure are observed twice (so 63 per cent of the sample). Thus for $m>24.6$, some galaxies might need a third observation (blue shade) to compensate for the gap between $E(m)$ and $p_{\rm min}$.
With our model, for efficiency $p_\text{min}<0.85$ it is never necessary to observe some object four times.

The average time dedicated to the observation of an object of magnitude $m$ (whether it is a success or not) is 
\begin{equation}
    \langle t(m) \rangle=\sum_n \eta_{n}(m) \cdot n\cdot t_\text{exp}.
    %(1\times \eta_\text{1obs}+2\times \eta_\text{2obs}(m)+\ldots),
    \label{tavg}
\end{equation}
The analytical expressions of $\eta_n$ and $\langle t(m) \rangle$ as function of $p_\text{min}$, $E(m)$ and  $m$ are given in Appendix \ref{sec:appendixA}.
\label{sec:average_time_and_efficiency}

\subsubsection{Survey duration and measured density}
\label{sec:surveyduration}
The observed volume is described by three numbers: $z_\text{min}$, $\Delta z$=$z_{\text{max}}-z_{\text{min}}$ and $f_\text{sky}$. From all available galaxies within this volume, we will visit a fraction $\eta_\text{til}  \lesssim1$ of them, and the visited tracer number is:
\begin{equation}
    N_\text{vis}=\eta_\text{til} \int_{z_\text{min}}^{z_\text{max}}dz \frac{d\chi}{dz}4\pi f_\text{sky}\chi^2 \int^{m_\text{max}}dm \phi_\text{LBG}(z,m).\label{eq:Nobs}
\end{equation}
This equation is the integration of the number density over the cosmic volume. The $\eta_\text{til}$ factor takes into account two effects: 
\begin{itemize}[wide, labelwidth=!,itemindent=!,labelindent=0pt, leftmargin=0em, itemsep=.1cm, parsep=0pt]
    \item The fibre collision: two close tracers cannot be observed simultaneously by two fibres during a single exposure,
    \item The tilling: the sky is generally not perfectly covered by the succession of focal planes, if objects are attributed to different exposures by maximising the survey efficiency, which is typically the case in reality.  
\end{itemize}
From this last equation, we further define the density of visited tracer per magnitude, 
\begin{equation}
    dN_\text{vis}(m)=4\pi f_\text{sky}\eta_\text{til} \int_{z_\text{min}}^{z_\text{max}}dz \frac{d\chi}{dz}\chi^2 \phi_\text{LBG}(z,m) dm.%\label{eq:Nobs}
\end{equation}
Given the average time to observe a tracer with magnitude $m$ $\langle t(m) \rangle$  (cf. equation \eqref{tavg}), and the fibre number $N_\text{fib}$, the total observational time of the survey is
\begin{equation}
\begin{aligned}    
\alpha T_\text{sur}&=\int^{m_\text{max}}dN_{\rm vis}(m)\langle t(m)\rangle/N_{\rm fib} \\
    &=\eta_\text{til}\frac{4\pi f_\text{sky} }{N_\text{fib}}\int_{z_\text{min}}^{z_\text{max}}dz \frac{d\chi}{dz}\chi^2 \int^{m_\text{max}}dm \langle t(m)\rangle\phi_\text{LBG}(m,z),
    \label{eq:eta_model}
\end{aligned}
\end{equation}
where $T_\text{sur}$ is the total survey duration (typically 5 years), and $\alpha $ is a coefficient to convert it into observational time. 
%$T_\text{sur}$ corresponds to the total amount of the observation time (an effective time). Thus for a 5-years survey, $T_\text{sur}=\alpha 5$ effective years, with $\alpha$ the ratio of observation time to total time. 
For cosmological observations, we assume we only observe days with a partial moon\footnote{Days with a full moon can be thus fully dedicated to other astrophysical observation.}, 21 days every 28 days cycle, within practice only 80 per cent of these nights that can be dedicated to observation (due to weather or maintenance issues)\footnote{For 5 years this corresponds to $\approx 1100$ nights.}, and between 8 and 10 hours of observation per night, which correspond to $\alpha=7\times10^6\,$yr$^{-1}$s. %During our discussion we omit this $\alpha$ factor since it is very , and we will discuss in term of the total time of the survey rather than the effective time.
In equation \eqref{eq:eta_model}  we are assuming that every fibre is dedicated to observation, and will be observed during every exposure. 

The density of observed tracers with a good spectroscopic redshift is
\begin{equation}
    n_\text{obs}(z)=\eta_\text{til}\int^{m_\text{max}}dm p(m)\phi_\text{LBG}(m,z).
    \label{eq:model_density}
\end{equation}
%Let us do few comments here. 
Since $T_\text{sur}\propto 1/N_\text{fib}$, the observation  for a 5 years survey with 20k fibre, would be equivalent to that of a 10-year survey with 10k fibres. Thus we introduce the `power'  parameter $N_\text{fib}\,T_\text{sur}$ in fibre-year.

\subsubsection{Optimisation Pipeline}
\label{subsub:strategy}
In the first part of our general survey study, we fix the following parameters: 
\begin{itemize}[wide, labelwidth=!,itemindent=!,labelindent=0pt, leftmargin=0em, itemsep=.1cm, parsep=0pt]
    \item $N_\text{fib}\,T_\text{sur}=100,000$ in fibre-year; it corresponds to 5 years of observation with 20,000 fibres for example;
    \item $\eta_\text{til}=0.96$, is the fraction of available tracer in our cosmic volume that we will observe; 
    \item $t_\text{exp}=1800\,s$, is the exposure time of the instrument;
    \item $p_\text{min}=0.7$, is the minimal efficiency rate;
    \item $z_\text{min}=2$, is the minimal redshift of our high-redshift surveys;
    \item $\alpha=7\times10^6$, as explain in section \ref{sec:surveyduration};
    \item The telescope size to be 10m;
\end{itemize}
and find the optimal survey volume described by:
\begin{itemize}[wide, labelwidth=!,itemindent=!,labelindent=0pt, leftmargin=0em, itemsep=.1cm, parsep=0pt]
    \item $f_\text{sky}$, the observed fraction of the sky, 
    \item $\Delta z=z_\text{max}-z_\text{min}$, the redshift width.
\end{itemize}
For arbitrary values of these two parameters, the equations \eqref{eq:eta_model} and \eqref{eq:model_density} will constrain:
\begin{itemize}[wide, labelwidth=!,itemindent=!,labelindent=0pt, leftmargin=0em, itemsep=.1cm, parsep=0pt]
    \item $m_\text{max}$ the maximal magnitude of observation,
    \item $n_\text{obs}(z)$ the tracer density,
\end{itemize}
according to the following scheme,
\begin{equation}
    f_\text{sky},\Delta z\overset{\text{Eq.~\eqref{eq:eta_model}}}{\longrightarrow}m_\text{max}\overset{\text{Eq. } \eqref{eq:model_density}}{\longrightarrow}n_\text{obs}(z)\overset{\mathcal{F}_{ij}}{\longrightarrow}\sigma(f_\text{NL}),\sigma(M_\nu).
    \label{eq:scheme}
\end{equation}
The numerical procedure to get $m_\text{max}$ is explained in Appendix \ref{sec:appendixB}.\\

In the second step, we fix the optimal survey volume, and free $N_\text{fib}\,T_\text{sur}$ and $p_\text{min}$ in a similar procedure as previously, mainly for two reasons. Firstly, it will show the optimal strategy between correctly measuring most of the objects visited (high $p_\text{min}$), or having a wider magnitude range and observing fainter objects that are located in the high redshift region. 
Secondly, it will quantify the improvement of the data over the fibre number, and the duration of the survey. We will vary $N_\text{fib}\,T_\text{sur}$ from 60,000 to 220,000 fibre-years.

\subsubsection{Observed Density Prediction}
\begin{figure}
    \centering
    \includegraphics[scale=0.6]{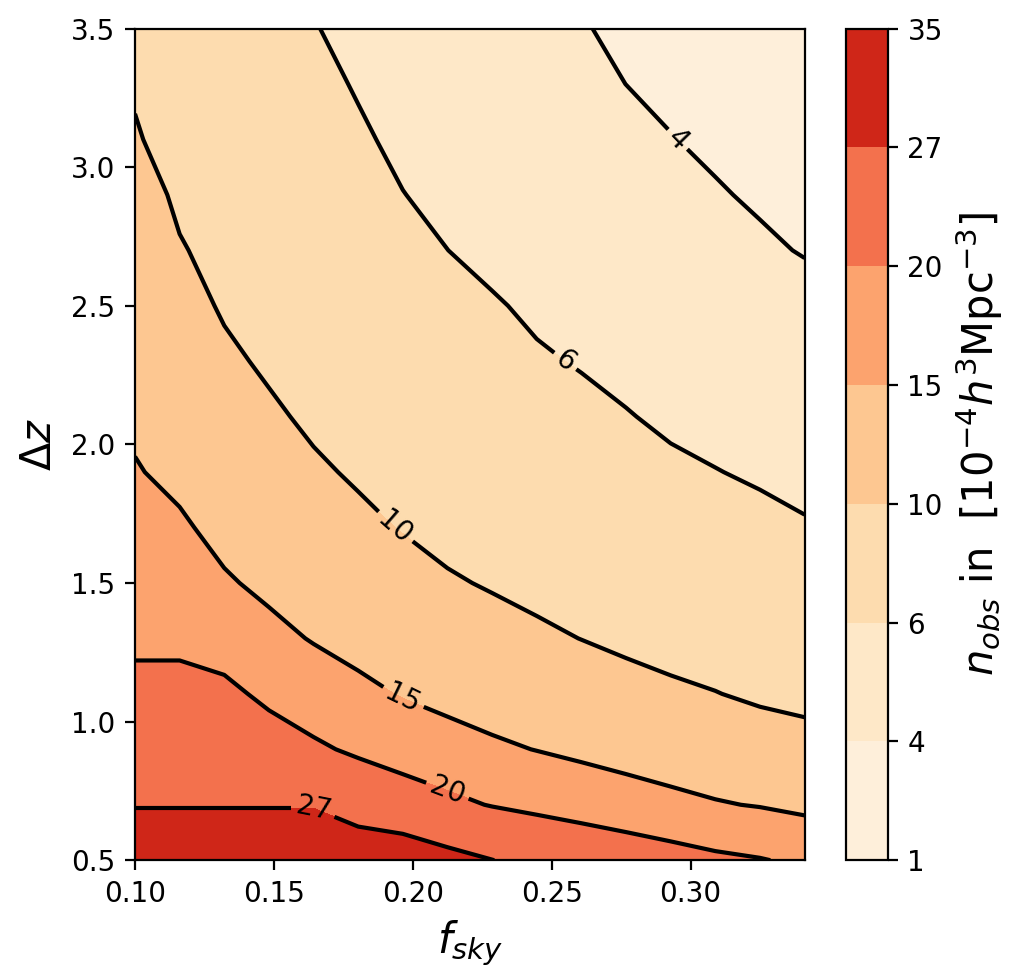}
    \caption{Observed density $n_{\rm obs}$ as a function of the redshift width $\Delta z$ and the sky coverage $f_{\rm sky}$, for $N_\text{fib}\,T_\text{sur}=100,000$ fibre-years, $z_{\rm min}=2$ and $p_\text{min}=0.7$. We observe hyperbolic trends, with saturation in the small volume region (bottom left corner), visible with the 20 and 27 $\times 10^{-4}h^3$Mpc$^{-3}$ lines.}
    \label{fig:density_dz_fsky}
\end{figure}

Before moving on to the cosmological parameters we show here some results of our modelling. We plot Figure~\ref{fig:density_dz_fsky}, the measured tracer density $n_{\rm obs}$ as a function of the survey cosmic volume, with parameters described in Section~\ref{subsub:strategy} (first two steps of equation \eqref{eq:scheme}).

\begin{figure}
    \centering
    \includegraphics[scale=0.6]{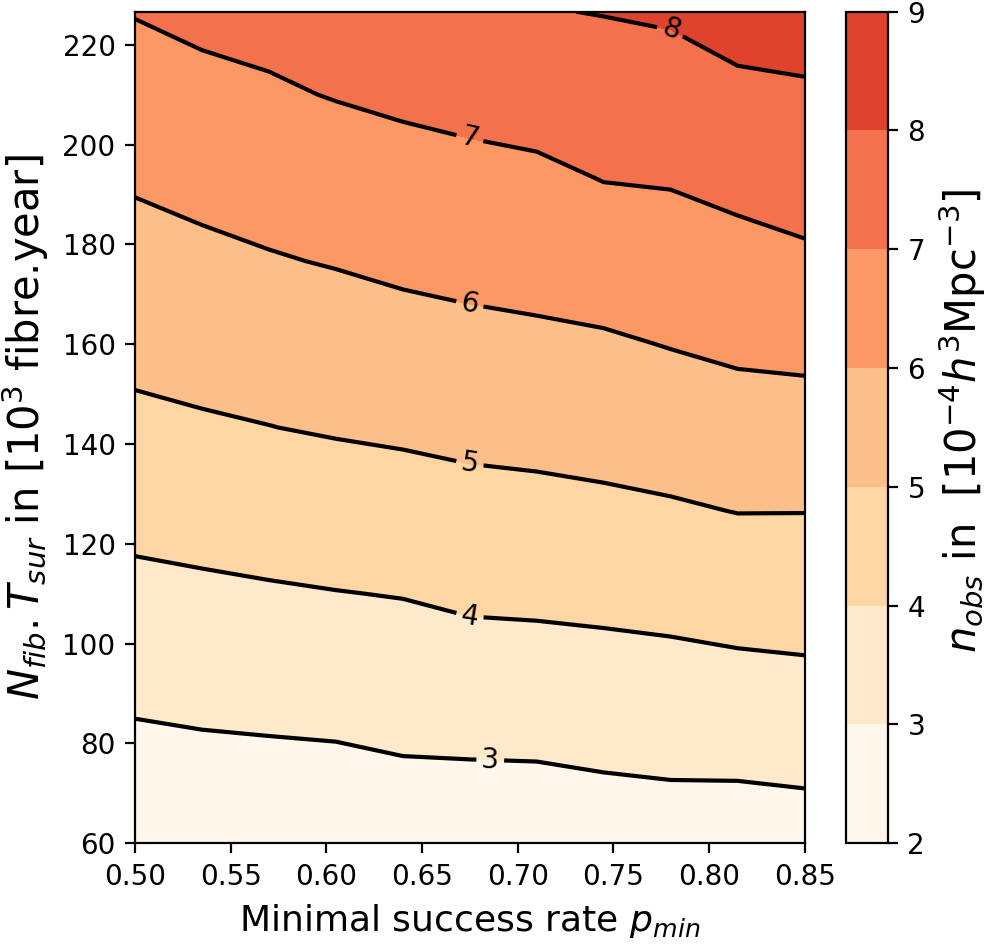}
    \caption{Observed density $n_{\rm obs}$ as a function of  $N_{\rm fib}T_{\rm sur}$, and minimal efficiency $p_{\rm min}$, for a maximal cosmic volume: $f_{\rm sky}=0.31$, $\Delta z=3$,and $z_{\rm min}=2$). The density increases linearly as a function of $N_{\rm fib}. T_{\rm sur}$, and increases with the efficiency threshold.}
    \label{fig:density_tsur_pc}
\end{figure}

We observe hyperbolic trends, as expected since the product of the two variables scales as the volume at first order (naive model). For small volumes ( e.g. $f_{\rm sky }=0.15$ and $\Delta z=0.7$), the maximum magnitude of 25 is reached (as illustrated by horizontal lines), and all tracers are visited by the end of the 5th year, which leads to saturation (cf Appendix \ref{sec:appendixB}).  
We also plot in Figure \ref{fig:density_tsur_pc} the average density as a function of $N_\text{fib}\,T_\text{sur}$, and the minimal efficiency $p_{\rm min}$. 
For a fixed $N_\text{fib}\,T_\text{sur}$, the density of observed tracers $n_{\rm obs}$ will be higher with higher minimum efficiency $p_{\rm min}$. Indeed if an observation of a magnitude $m_1$ object failed in the first exposure, then its second-exposure-observation is more likely to be successful, than a first-exposure-observation of another object of magnitude $m_2\geq m_1$\footnote{Indeed, during the second exposure, the initial SNR value will be the one obtained at the end of the first exposure.}. Nonetheless, with lower minimal efficiency, the maximal magnitude is higher and one is observing more objects at higher redshift.
Thus it is a priori difficult to know which strategy will provide the best constraints on the cosmological parameters.\\
%For large volume studies $f_\text{sky}\sim 0.3$ and $\Delta z=3$, we have $n_\text{meas}\sim 3.5\times 10^{-4}h^3/$Mpc$^3$.\\

With this general survey model, one should be able to reproduce fiducial tracer properties of future surveys with known observational parameters such as MegaMapper. It is a large volume survey with $f_\text{sky}\sim 0.3$ and $\Delta z=3$, but with a smaller telescope diameter ($\sim 6.5$ m) than the one assumed for $E(m)$ cf. section \ref{sec:one_exp}. Motivated by the dependence of the CCD equation \eqref{eq:CCD} on $\sqrt{ S_{\rm tel}}$, we correct the efficiency law $E(m)$ by a factor $\sqrt{S_\text{Mega}/S_\text{MSE}}$ and require a minimum efficiency $p_\text{min}=0.5$ (Section~\ref{sec:mega}). The predicted number density is $n_\text{obs}\sim 2.6\times  10^{-4}\,h^3\,\rm{Mpc}^{-3}$, very close to the expectation of MegaMapper which is $2.5 \times  10^{-4}\,h^3\,\rm{Mpc}^{-3}$. The maximum magnitude imposed by the model is 24.6, which is in agreement with MegaMapper's 24.5 maximal magnitude. It should be noted that this agreement is remarkable since our modelling is independent of any MegaMapper settings.

\section{Results}
\label{sec:results}
In this section, we first present the  BAO, RSD, NG, and NM forecasts for future surveys such as MUST, MegaMapper, and MSE. We then investigate the NM and NG accuracy given the survey properties with our model introduced in Section~\ref{sec:GS}. We deduce, independently of any planned survey, the optimal observation strategies, and the limits when measuring these parameters.

\subsection{BAO and RSD Constraints}
\label{subsec:bao rsd constraints}
We summarised in Table~\ref{tab:RSDBAO} the constraints on BAO and RSD parameters. We separate MegaMapper and MSE redshift range into several bins and present the forecast for every bin. We also provide constraints in the full redshift range in the last row for each survey. We derived the constraints for 8 different settings of MUST, and for 3 different settings of an NTL survey, that mimics the observation of the WST survey (depending on the fibre number). Finally, we consider a  combination of two MegaMapper-like surveys, each located in one hemisphere. This study shows the power of combined independent surveys in providing better cosmological constraints.

\begin{table*}
\centering
\caption{The predicted 68 per cent Confidence Level (CL) error of the BAO distances and RSD parameters for various surveys. We use separate redshift bins for MegaMapper and MSE, and we also show the forecast using tracers at the whole redshift range in the last row of each survey. We present the forecast for MUST at $2<z<4$ and for NTL and combined surveys at $2<z<5$. LBGX$\times$LAE denotes a multi-tracer constraint with LBGX and LAE.}
\begin{tabular}{cccccccccc}
\hline
\multicolumn{10}{c}{BAO and RSD forecast} \\
\hline
\makecell{Fibre \\ number} &
\makecell{Sky area \\ deg$^2$} &
Tracer &
Redshift &
\makecell{Number density \\ $10^{-4}\,h^3\,{\rm Mpc}^{-3}$} &
$n P(0.14, 0.6)$ &
\makecell{$\sigma(D_A)/D_A$ \\ (\%)} &
\makecell{$\sigma(H)/H$ \\ (\%)} &
\makecell{$\sigma(b\sigma_8)/b\sigma_8$ \\ (\%)} &
\makecell{$\sigma(f\sigma_8)/f\sigma_8$ \\ (\%)}\\
\hline
\multicolumn{10}{c}{\textbf{MegaMapper}} \\
\multirow{5}{*}{20k} & \multirow{5}{*}{14k} & \multirow{5}{*}{LBG} & $2<z<2.5$    & 7.9&1.7 &0.32 &1.0&0.066 &0.85\\
&&& $2.5<z<3$   &3.6 &0.68 &0.35 & 1.0&0.073 &1.0\\
&&&$3<z<4$ & 1.1& 0.19& 0.40& 1.0&0.085&1.3\\
&&&$4<z<5$ &0.7 &0.11 & 0.45&0.99&0.094 &1.7\\
&&&$2<z<5$ &2.5 &0.5 & 0.18&0.57&0.039 &0.52\\
\hline
\multicolumn{10}{c}{\textbf{MSE}} \\
\multirow{5}{*}{4.3k} & \multirow{5}{*}{10k} & ELG & $1.6<z<2.4$    &1.8& 0.28&0.86 &2.5 &0.17 & 1.4\\
 &&LBG& $2.4<z<2.8$   &2.3&0.51&0.51&1.5&0.10&1.7\\
 &&LBG& $2.8<z<3.2$   &1.1&0.22&0.66&1.8&0.14&2.3\\
 &&LBG&$3.2<z<4$ & 0.43& 0.08&0.79&1.9&0.17&2.8\\
 &&LBG&$2.4<z<4$ &1.1&0.28&0.28&0.64&0.078&1.2 \\
\hline
\multicolumn{10}{c}{\textbf{MUST (different settings)}} \\
 20k&15k& LBGX&$2<z<4$&8.9&2.3&0.13&0.41&0.026&0.44\\
20k&15k&LBGX$\times$LAE&$2<z<4$&4.2--0.84&1.1&0.15&0.49&0.030--0.14&0.47\\
10k&15k&LBGX&$2<z<4$&5.0&1.3&0.15&0.48&0.030&0.52\\
10k&15k&LBGX$\times$LAE&$2<z<4$&2.1--0.42&0.58&0.19&0.62&0.039--0.21&0.60\\
%5k&15k&LBGX&$2<z<4$&2.5&0.64&0.18&0.60&0.038&0.64\\
%5k&15k&LBGX$\times$LAE&$2<z<4$&1.1--0.25&0.30&0.26&0.82&0.052--0.31&0.079\\
10k&9k&LBG&$2<z<4$&7.1&2.1&0.18&0.56&0.035&0.60\\
10k&9k&LBGX$\times$LAE&$2<z<4$&3.5--0.71&0.97&0.21&0.68&0.041--0.21&0.66\\
\hline
\multicolumn{10}{c}{\textbf{NTL (WST like survey)}} \\
 20k&15k& LBG& $2<z<5$    & 2.5&0.51 & 0.15& 0.48&0.035&0.54\\
 40k&15k&LBG& $2<z<5$    & 4.9&0.99 & 0.12& 0.38&0.030&0.40\\
 100k&15k&LBG &$2<z<5$   &13 &1.9 & 0.099& 0.29&0.027&0.26\\
\hline
\multicolumn{10}{c}{\textbf{Combination of two MegaMapper-like surveys}}\\
  20k& 28k&LBG&$2<z<5$& 2.5&0.5&0.13&0.40&0.028&0.37\\
\hline
\end{tabular}
\label{tab:RSDBAO}
\end{table*}

From this table we can conclude that:
\begin{enumerate}[wide, labelwidth=!,itemindent=!,labelindent=0pt, leftmargin=0em, itemsep=.1cm, parsep=0pt,label=(\roman*)]
 \item MegaMapper and MUST (20,000 fibres) have relatively similar accuracy. In theory, they will improve the constraints on BAO and RSD by a factor of 10 w.r.t. eBOSS constraints \citep{eBOSSforecast}, and a factor between 2 and 3 w.r.t. DESI constraints \citep{desi}. MSE forecasts are not as good as these two because of its smaller fibre number. 
\item For MUST, with the same number of fibre, the one-tracer (LBGX) case gives better constraints than the two tracers (LBGX+LAE) one. It is mainly because LAEs need a long exposure time, which decreases the total number of observed LAEs and thus increases the overall noise, as illustrated by the $nP$ parameter values.
\item The combination of two independent LBG surveys gives cosmological constraints similar to those of an NTL survey with 40,000 fibres, a factor of $\sqrt{2}$ smaller than those of a single survey like MegaMapper. As the two surveys are independent, this corresponds to the constraints on the BAO and RSD from two independent measurements, combined.
\item 100,000 fibres in the NTL survey represent the upper limit for these high-redshift LBG surveys since it corresponds to the measurement of all galaxies up to $m_\text{max}=25$.
\item All the parameters are very well constrained down to the sub-per-cent level in every survey. This is unrealistic as the real observational constraints are likely to be dominated by systematics, and not anymore by  the number density of tracers and the  cosmic volume.
\end{enumerate}
As the sample size  will not bring improvement for cosmological measurements of future surveys, we do not optimise it in the following study. 

\subsection{Non-Gaussianity and Neutrino Mass}
\subsubsection{Redshift Binning}
As mentioned in Section~\ref{sec:redshift_binnung}, there are several ways to deal with the large cosmic volume. In Section~\ref{subsec:bao rsd constraints}, as their parameters are redshift-dependent, it is natural to provide forecasts in small redshift bins. In contrast, non-Gaussianity and neutrino mass are independent of the redshift. Thus splitting the samples into multiple redshift bins does not bring more information for their measurements in principle. However, as the number density of tracers is much higher at low redshift than that at high redshift, analysing  the total redshift range is not necessarily the best option. Because it may overestimate the noise that scales as $1/n$ at low redshift.  

\begin{figure*}
    \centering
    \includegraphics[scale=0.6]{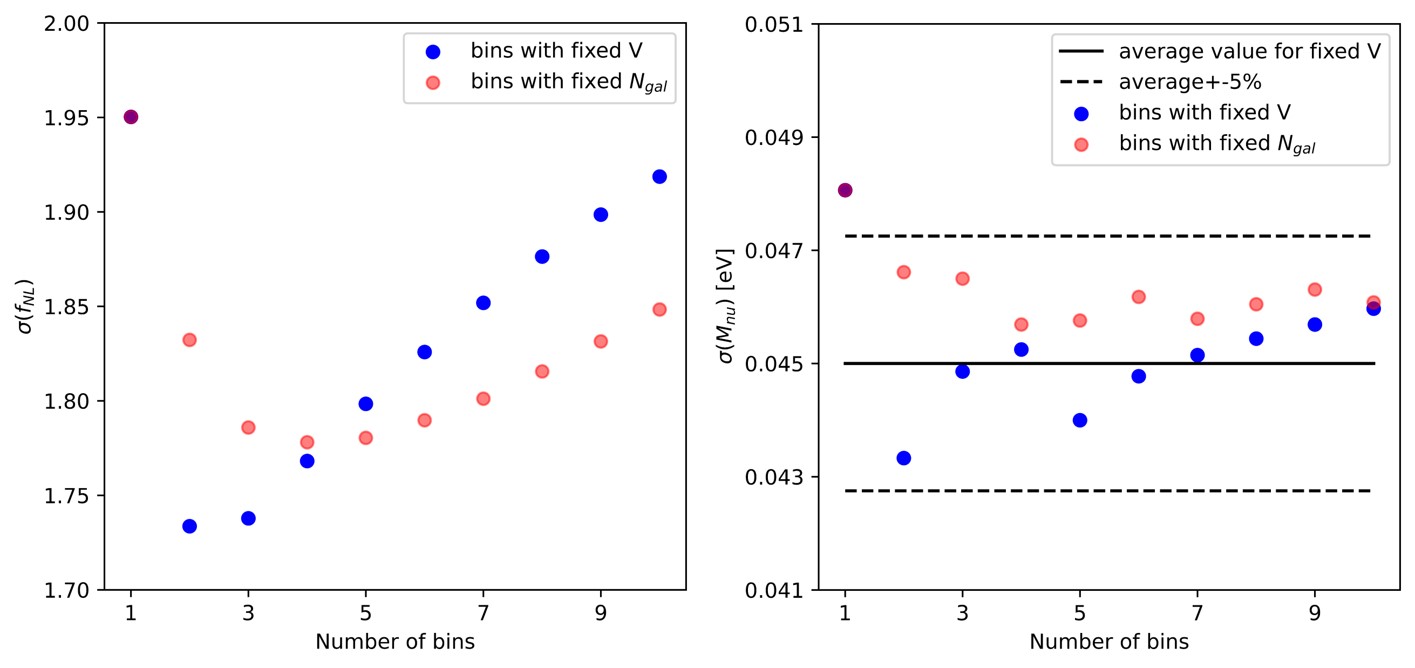}
    \caption{NM (the left panel) and NG (the right panel) forecast w.r.t. the bin number, for two binning schemes ($V$-based in blue dots and $N_{\rm gal}$-based in pink dots). For NM we also include the average value for the fixed volume approach (solid lines) and delimit the $\pm 5 $ per cent region around it by the two dashed lines. For NG, the approach with fixed volume is clearly the optimal one with 2 or 3 bins, whereas, for NM, there is not the best choice.}
    \label{fig:bin_NG_Nm}
\end{figure*}

We, therefore, investigate the optimal binning, by exploring two ways to separate our redshift interval $[z_\text{min},z_\text{max}]$ into $N_\text{bins}$:
\begin{itemize}[wide, labelwidth=!,itemindent=!,labelindent=0pt, leftmargin=0em, itemsep=.1cm, parsep=0pt]
    \item define bins with the same comoving volume $V$: a `fixed volume approach';
    \item define bins with the same number of targets $N_{\rm gal}$. In this case, the low-redshift bins have volumes smaller than those at higher-redshift bins.  
\end{itemize}
A model including a continuous dependence of the redshift and density is more promising, but we leave this more complex option for a future study. 
We calculate $\sigma(f_\text{NL})$ and $\sigma(M_\nu)$ for $N_\text{bins}=1,2,...,10$, and report their relation in Figure \ref{fig:bin_NG_Nm}. We fix the tracer density and cosmic volume for these forecasts to be the fiducial MegaMapper ones.
We check that the optimal binning scheme is independent of these parameters.

For non-Gaussianity, the fixed volume approach provides the best constraints with 2 or 3 bins. Indeed $f_\text{NL}$ describes a large-scale phenomenon, and reducing bin sizes increases the low integration limit value $k_\text{min}$, which leads to a rapid increase in variances for a large number of bins. The drop in $\fnl$ accuracy from one bin to two bins is due to the reduction in noise at the low redshift bin which compensates for its small volume\footnote{This difference is smaller in the case of the fixed number of tracer. Because in that case, the volume of the low-redshift bin is small, and we have a small bin with low noise, and a large bin with high noise at high redshift.}.

For neutrino mass measurement, contrary to $\fnl$, there is no clear pattern, and therefore no number of bins has to be preferred. In fact, except for one bin, the results fluctuate by less than $5$ per cent from the mean value, which is not large given our method. This is a confirmation that this forecast does not depend on large scales, but on small ones as theoretically expected.

In the following studies, we adopt a volume-fixed approach with 3 bins for both forecasts. 

\subsubsection{Small scale choice}

We do the forecast with two different $k_\text{max}$ values: 0.1 and 0.3 $h\rm{Mpc}^{-1}$. The $k_{\rm max}$ reached in the future survey analysis may be between these two, or even larger. However, for $k_\text{max}=0.3$ $h\rm{Mpc}^{-1}$, our linear approach is already insufficient, and a proper future analysis would have to take into account non-linear correction as what \citet{white_same, neutrino_nonlinear} have done. This will mainly affect the neutrino forecast since non-Gaussianity is a large-scale phenomenon, whereas the power spectrum is more affected by the massive neutrino contribution at large $k$. %It is highlighted in our table since the $\sigma(f_{\text{NL}})$ is affected by less than 10$\%$ by the $k_\text{max}$ increase, whereas $\sigma(M_\nu)$ may change by a factor of 2. 

\subsubsection{Forecasts}
In Table~\ref{tab:NGNM} we present forecasts for the same surveys as Table~\ref{tab:RSDBAO}, but for the neutrino mass and non-Gaussianity. For the neutrino forecast, we provide constraints with only a prior from Planck CMB, and those with a Planck CMB prior and measurements from DESI, as described in Section~\ref{sec:planck_desi}. 

\begin{table*}
\centering
\caption{The same Table as as Table~\ref{tab:RSDBAO} but for $f_\text{NL}$ and $M_\nu$. The forecast is based on a high mode integration limit $k_\text{max}=0.1-0.3~ h\rm{Mpc}^{-1}$, which corresponds to a pessimistic and an optimistic scenario (that is why we provide two values for each parameter). The left column for $M_\nu$ is the forecast with a CMB prior, whereas the right column is the result of a combination of DESI and a CMB prior.}
\begin{tabular}{cccccccc}
\hline
\multicolumn{8}{c}{Non Gaussianity and neutrino mass forecast for $k_{\text{max}}=0.1$--$0.3~h$Mpc$^{-1}$} \\
\hline
\makecell{Fibre \\ number} &
\makecell{Sky area \\ deg$^2$} &
Tracer &
Redshift &
\makecell{Number density \\ $10^{-4}\,h^3\,{\rm Mpc}^{-3}$} &
\makecell{$\sigma(\fnl)$} &
\makecell{$\sigma(M_\nu)$ in $10^{-2} \text{eV}$\\ Planck} &
\makecell{$\sigma(M_\nu)$ in  $10^{-2} \text{eV}$\\ Planck and DESI}\\
\hline
\multicolumn{8}{c}{\textbf{MegaMapper}} \\
20k&14k&LBG&$2<z<5$ &2.5 & 1.3--1.2&4.2--2.4&3.3--1.6 \\
\hline
\multicolumn{8}{c}{\textbf{MSE}} \\
\multirow{2}{*}{4.3k}  & \multirow{2}{*}{10k}  & ELG& $1.6<z<2.4$   &1.8&8.9--8.1&5.0--4.5&4.1--3.0\\
& & LBG& $2.4<z<4$   &1.1&2.8--2.5&5.1--4.3&3.7--2.1\\
\hline
\multicolumn{8}{c}{\textbf{MUST (different settings)}} \\
 20k&15k& LBGX&$2<z<4$&8.9&1.6--1.4&4.2--2.4&3.1--1.6\\
20k&15k&LBGX$\times$LAE&$2<z<4$&4.2--0.84&1.7--1.5&4.4--2.9&3.2--1.8\\
10k&15k&LBGX&$2<z<4$&5.0&1.7--1.5&4.3--2.8&3.2--1.7\\
10k&15k&LBGX$\times$LAE&$2<z<4$&2.1--0.42&1.9--1.8&4.5--3.4&3.3--2.0\\
%5k&15k&LBGX&$2<z<4$&2.5&1.8--1.6&4.5--3.3&3.7--2.0\\
%5k&15k&LBGX$\times$LAE&$2<z<4$&1.1--0.25&2.1--2.0&4.7--3.8&3.8--2.3\\
10k&9k&LBG&$2<z<4$&7.1&2.1--1.8&4.5--2.8&3.2--1.8\\
10k&9k&LBGX$\times$LAE&$2<z<4$&3.5--0.71&2.2--2.0&4.6--3.4&3.3--2.0\\
\hline
\multicolumn{8}{c}{\textbf{NTL (WST like survey)}} \\
 20k&15k& LBG& $2<z<5$    & 2.5 &1.4--1.4&4.1--2.1&3.3--1.5\\
 40k&15k&LBG& $2<z<5$    & 4.9 & 1.1--1.0&3.6--1.7&3.0--1.3\\
 100k&15k&LBG &$2<z<5$   &13  &1.0--0.90&3.2--1.3&2.7--1.0\\
\hline
\multicolumn{8}{c}{\textbf{Combination of two MegaMapper-like surveys}}\\
  20k& 28k&LBG&$2<z<5$& 2.5&0.92--0.85&3.7--1.8&3.0--1.3\\
\hline
\end{tabular}
  \label{tab:NGNM}
\end{table*}
For non-Gaussianity, we may reach with MegaMapper-like survey $\sigma(f_\text{NL})\sim 1.2$, which is a factor of 4 better than the DESI forecast \citep{desi}. In particular, the constraints depend mainly on the volume of the survey. For example, with the number of fibres reduced by a factor 2 (which propagates to the tracer density), constraints of MUST vary from 1.6 to 1.7 for $k_\text{max}=0.1h$Mpc$^{-1}$; whereas reducing the observation window from 15,000 to 9,000 deg$^2$ leads to a weaker constraint from 1.7 to 2.1 for 10,000 fibres (despite a higher tracer density for the reduced window).  
$k_\text{max}$ values have little impact on $f_\text{NL}$, which is within expectation since it describes a large scale deviation from classical Gaussian prediction.  A combination of two surveys (coverage of 28,000 square degrees) results in a smaller $\sigma(\fnl)$, than the NTL with 100,000 fibres, which shows that the measurement of this parameter is limited by the volume and not the tracer density. That is also why, MSE constraints are much weaker than the ones of the other surveys.

\begin{figure}
    \centering
    \includegraphics[scale=0.6]{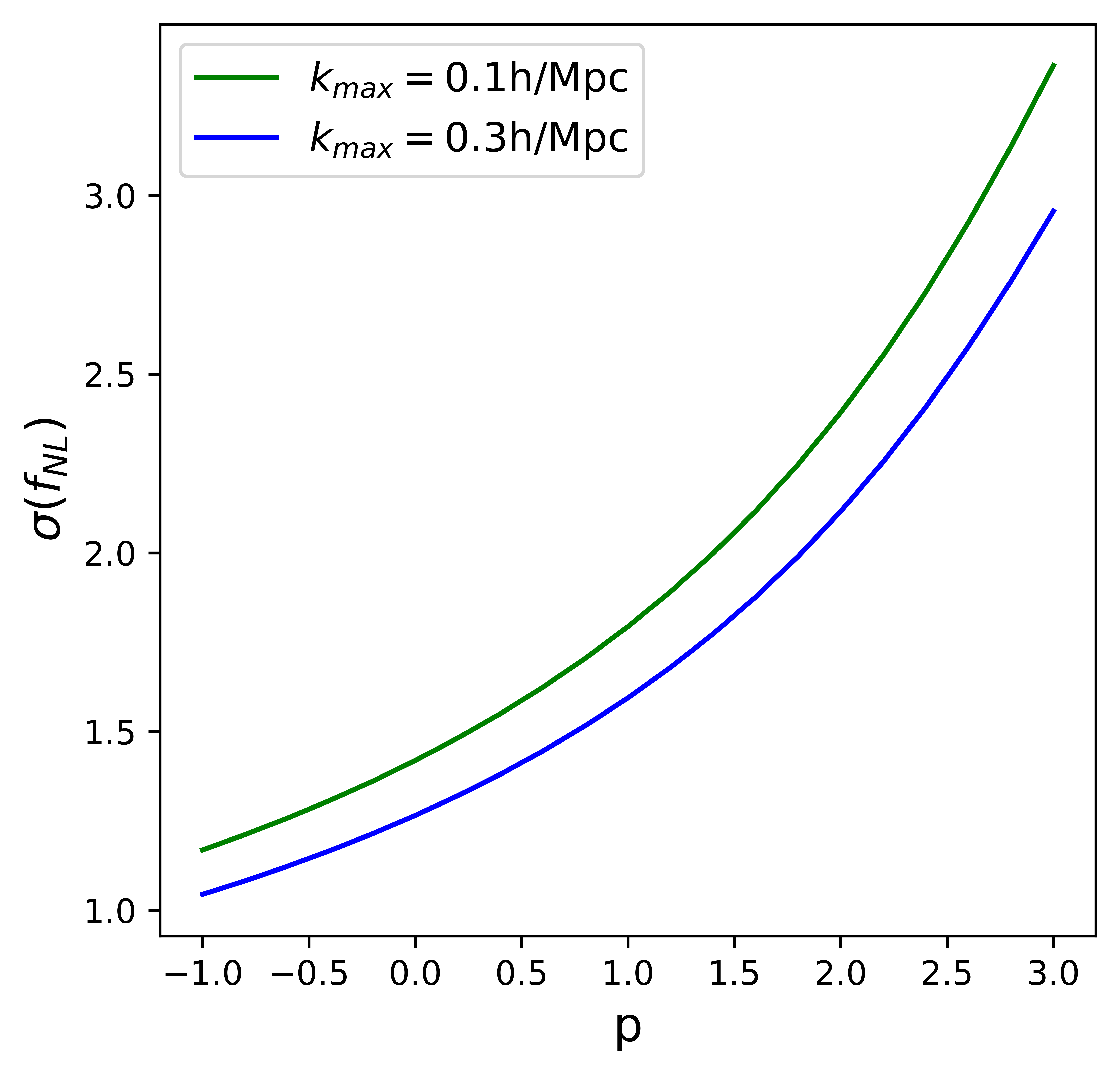}
    \caption{The predicted constraints on non-Gaussianity for a high-redshift LBG survey  w.r.t. the fiducial p value.}
    \label{fig:fnl_p}
\end{figure}

As already mentioned in Section \ref{sec:fnl_theory}, we assume a value $p=1$, but indeed different values are theoretically possible. The NG amplitude is degenerated with $p$, as it scales as $f_\text{NL}p$, as well as the variance We illustrate this via the $\sigma(f_\text{NL})$ forecast for a MegaMapper-like survey as a function of the $p$ values in Figure~\ref{fig:fnl_p}. In particular, for relatively reasonable values like $p=1.6$, the variance on NG amplitude is already significantly degraded ($\sim30$ per cent w.r.t. $p=1$). Thus, in order to have a reliable measurement of the $\fnl$ parameter (value and uncertainty), one needs robust theoretical priors, provided, for example, by simulations. One can find a complete discussion about this issue in \citet{fnl_and_p}.

The neutrino forecast depends strongly on the value of the small scale limit $k_\text{max}$ (by a factor of two in most cases). With a CMB prior, $\sigma(M_\nu)$ values are relatively similar for most surveys of the order of 0.04eV for $k_\text{max}=0.1$, and 0.025eV for $k_\text{max}=0.3$. With our approach, we find an improvement of about 50 per cent w.r.t. CMB+DESI forecast only. 
Combining both, it is possible to reach 0.015 eV. However forecasting this parameter with our naive approach aims to show global trends, and values should not be taken in the strict sense for the reasons mentioned in section \ref{sec:neutrino_limit}, but rather as indicators. Unlike the NG, the NM measurement of the combination of two 
surveys is worse than the NTL 100,000 fibres. Thus this parameter is more impacted by noise. 

\begin{figure}
    \centering
    \includegraphics[scale=0.36]{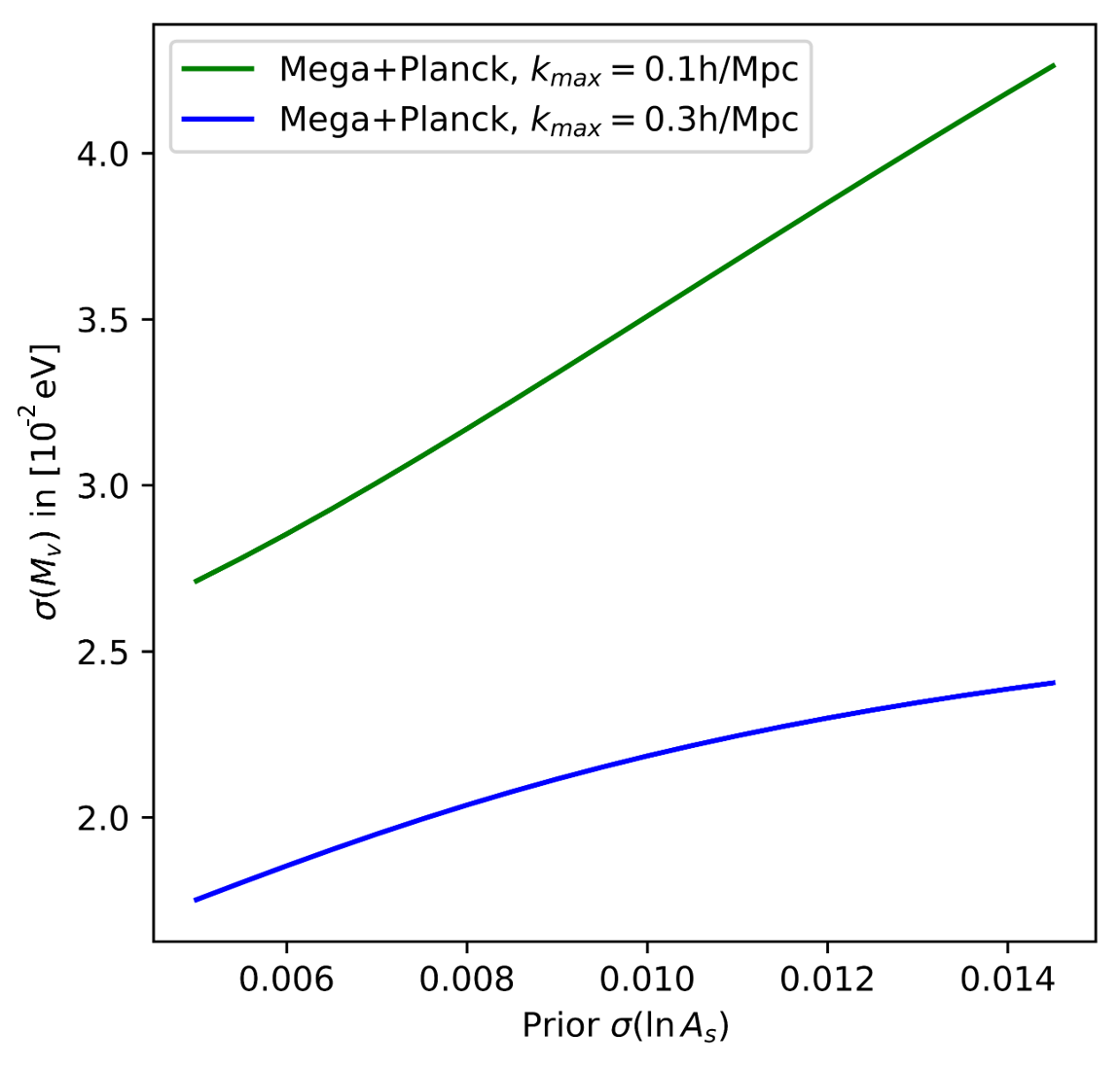}
    \caption{The predicted neutrino mass constraints for a MegaMapper-like survey w.r.t. the $\ln{A_s}$ prior. For the other cosmological parameters, prior values are taken from Planck CMB measurements.}
    \label{fig:mnu_prior}
\end{figure}

The neutrino forecast varies slightly with the characteristics of the survey. Indeed, the limiting factor for the accuracy of this parameter is the weak prior of some parameters from the standard model, rather than the survey settings. As discussed in \citet{thesis_neutrino,neutrino_nonlinear}, it is due to our limited knowledge of $A_s$, which is linked to $\tau$ because of a strong CMB degeneracy between these two parameters \footnote{Indeed $A_s$ is not directly constrained but $\tau$ and $A_s\exp(-2\tau)$ are.}. Table 2 of \citet{21cm_cosmo} summarises forecasts of the expected future knowledge on $\ln(A_s)$ and $\tau$ with future 21cm surveys. We illustrate this issue in Figure~\ref{fig:mnu_prior}, by plotting $\sigma(M_\nu)$ for a MegaMapper-like survey, as a function of the prior on $\ln{A_s}$, with values credible according to Table 2 of \citet{21cm_cosmo}.  For example, a difference in the $\sigma(\ln{A_s})$ prior from 0.014 to 0.005 leads to an improvement from 0.024eV to 0.018eV, in the optimistic scenario. Thus improving the knowledge on $A_s$ is the best way to detect the neutrino mass hierarchy as large as a $5\sigma$ confidence level. For other cosmological parameters, we find that the dependencies are much weaker, and we simply take the CMB prior values.

\subsection{Optimised NG surveys}
\label{subsec:optimise NG}
We now study the optimized survey characteristics to minimise $\sigma(\fnl)$, with the survey model described in Section~\ref{sec:GS}. The fixed-volume redshift bin number is given by
\begin{equation}
   N_{\rm{bin}} =\left\{
\begin{array}{ll}
1 &\text{if } \Delta z<1 \\
2&\text{if } 1\leq \Delta z<2\\
3&\text{if } 2\leq \Delta z.
\end{array}
\right.
\end{equation}

\begin{figure*}
    \centering
    \includegraphics[scale=0.6]{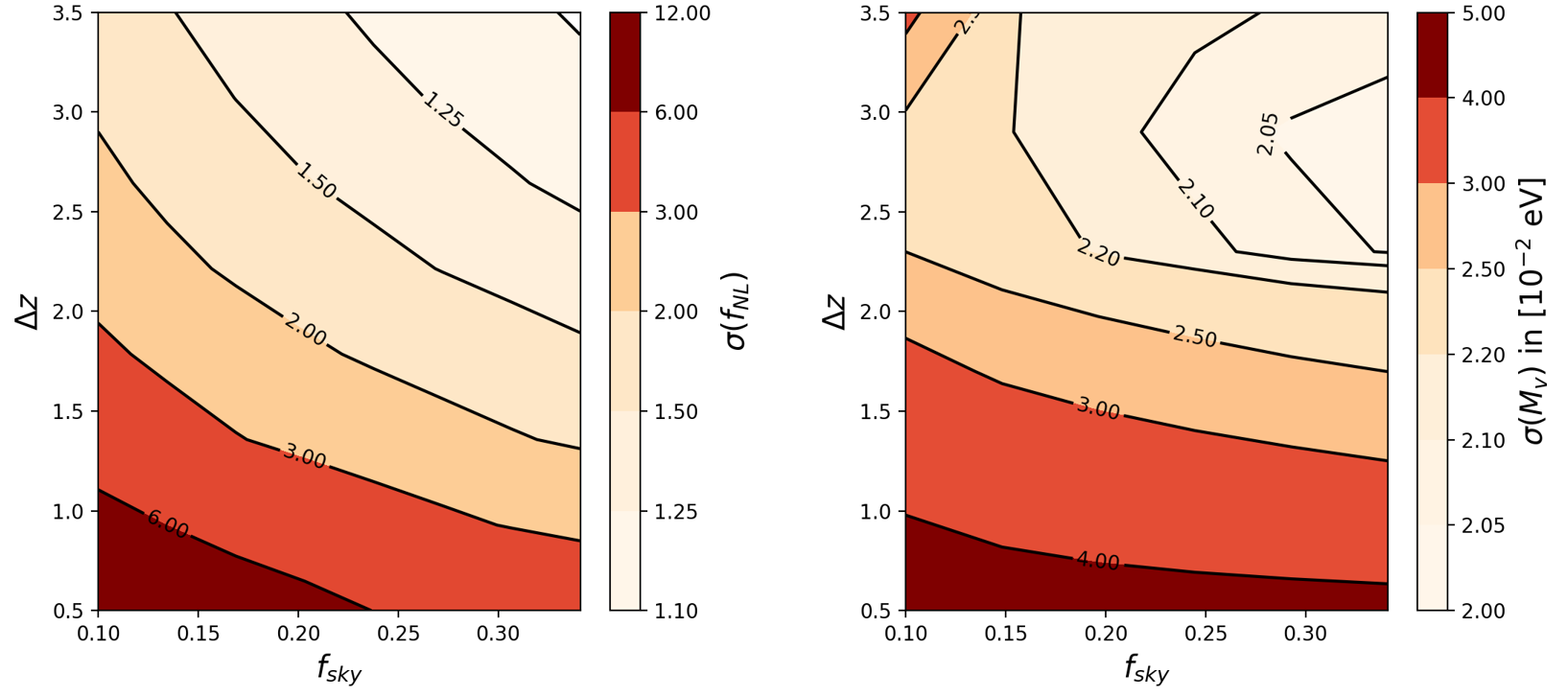}
    \caption{Non-Gaussianity (left panel) and massive neutrino (right panel) general survey forecasts as a function of the cosmic volume describes by the fraction of the sky $f_{\rm sky}$ and redshift width $\Delta z$, for 100,000 fibre-year survey,  $z_{\rm min}=2$ and $p_{\rm min}=0.7$.}
    \label{fig:fnl_dz_fsky}
\end{figure*}

We start this analysis by varying the cosmic volume in Figure~\ref{fig:fnl_dz_fsky}, for a survey with 100,000 fibre-years, and a minimum efficiency of 0.7.
The improvement of the precision goes hand in hand with the increase of the volume. Thus for the study of $\fnl$, one should always favour a larger survey volume. It is not beyond expectation since $\fnl$ is a parameter related to the large scale. In addition to that, it depended little on tracer density as we show in Table~\ref{tab:NGNM}. This remains true when varying the survey strategy parameters, such as the survey duration and the minimum efficiency.

\begin{figure*}
    \centering
    \includegraphics[scale=0.6]{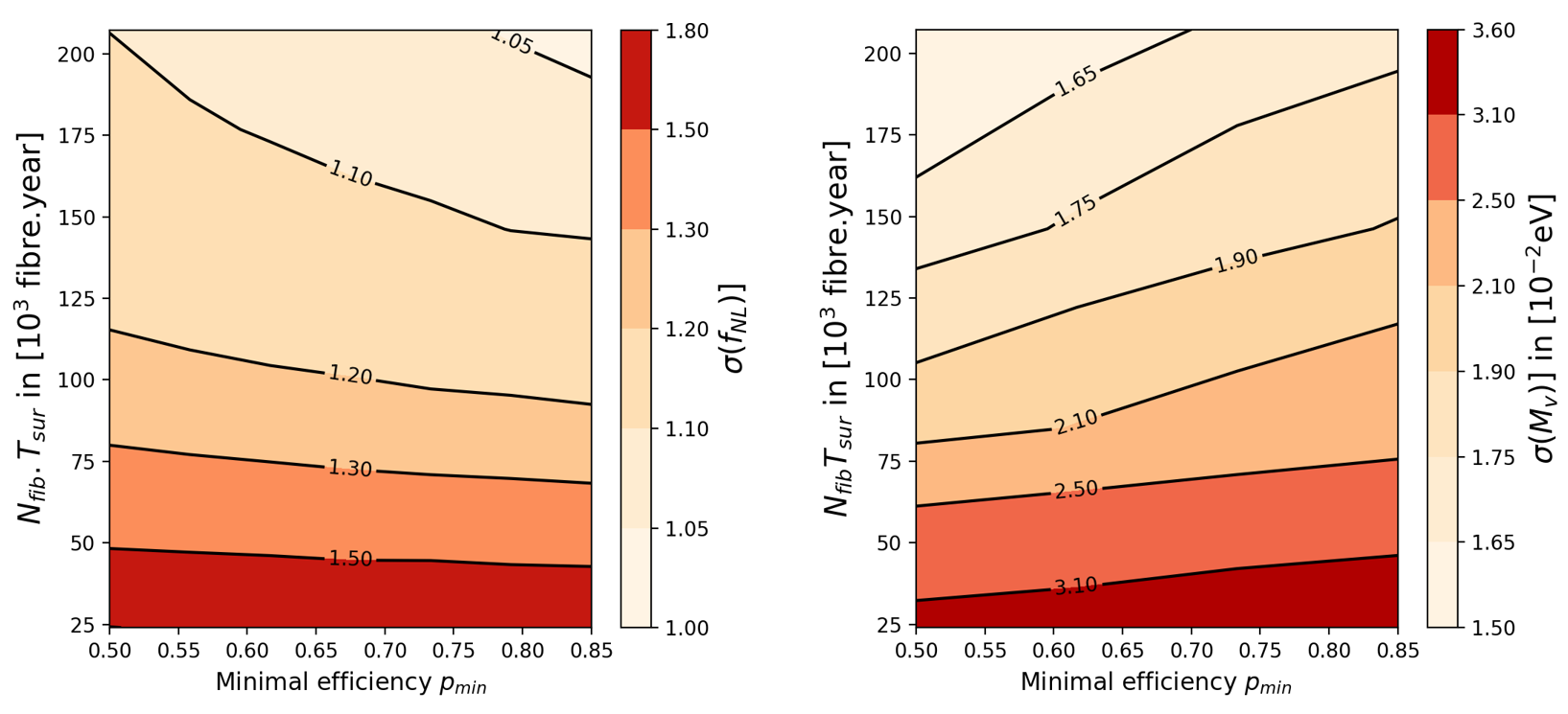}
    \caption{Non-Gaussianity (left panel) and massive neutrino (right panel) general survey forecasts as a function of the efficiency threshold $p_{\rm min}$ and $N_{\rm fib}. T_{\rm sur}$, for maximal cosmic volume: $f_{\rm sky}=0.31$, $\Delta z=3$, and $z_{\rm min}=2$.}
    \label{fig:fnl_tsur_pm}
\end{figure*}

Then we set a maximum volume among all the surveys and vary the fibre-year parameter $N_{\rm fib}T_{\rm sur}$, as well as the minimum efficiency $p_{\rm min}$. The forecast is reported in Figure~\ref{fig:fnl_tsur_pm}.
The NM accuracy depends weakly on the efficiency threshold but still deprives the high ones. Thus it is optimal to maximise the tracer average density and to restrict the maximum magnitude of observation, as it reduces the noise. However, high-redshift bins are populated with fainter galaxies and limiting the maximal magnitude also limits the tracer density at high redshift. This mutual restriction may explain why the tendency is weak. There is also a second contribution: we consider that the effective bias is the bias of tracers with the maximum magnitude  $b(z,m)\approx b_\text{LBG}(z,m_\text{max})$ (Section~\ref{sec:tracers}). Since bias decreases with magnitude, we underestimate the small-magnitude tracer bias. Since the NG effect scales as $\Delta b= f_\text{NL}(b-p)$, it also reduces the amplitude of the NG. 

Furthermore, multiplying the fibre-year parameter by four only increases the accuracy by 50 per cent. The main limitation neither comes from $N_\text{fib}\,T_\text{sur}$ nor $k_\text{max}$, but the cosmic volume of the survey $V_\text{sur}$. However, since with a longer survey time, one is able to go to a higher magnitude, it underestimates the variance of NG amplitude because the bias decreases as described previously. 

\subsection{Optimised NM surveys}
We adopt the same binning choice as for the NG study in Section~\ref{subsec:optimise NG}. We first show the results for different survey volumes in Figure 
%\ref{fig:mnu_dz_fsky}. 
\ref{fig:fnl_dz_fsky}. 
The variance of $M_\nu$ is less correlated with volume than for NG, especially for $f_\text{sky}$ between 0.2 and 0.3, and $\Delta z$ between 2.5 and 3.5. The optimal region seems to be around $\Delta z=3$ and $f_\text{sky}=0.3$, but the variation of $\sigma(M_\nu)$ in the nearby if these parameters is comparable with the accuracy of our naive model ($\sim 10$ per cent). 
Thus these results show a relatively good region, rather than a clear-cut result, with a trade-off between the volume and noise. Furthermore, the density of LBGs at high redshift is low, and including high redshift galaxies does not help constrain NM given the large shot noise, beyond $\Delta z=3$.

In accordance with the previously observed optimal region, we set the volume at $f_\text{sky}=0.31$ and $\Delta z=3$. In Figure~\ref{fig:fnl_tsur_pm}
we present the forecast varying the fibre-year parameter and the minimum efficiency. The NM results are quite different from those of NG. Indeed, it is clear that the NM constraint improves with an increasing fibre-year. The error is divided by 2 when the fibre-year parameter is multiplied by 4, which means that the shot noise level plays an important role. 
Moreover, the best is to work with a minimum efficiency, in order to observe objects with a low luminosity, located at high redshift. We have verified that for slightly different cosmic volumes close to the maximal one, these results are unchanged.

\section{Conclusion}
\label{sec:conclusion}
As successors of DESI, MegaMapper, MSE, and MUST will become the largest spectroscopic surveys in the world in the next decades. They will map the Universe in the redshift range $2 < z< 5$ using multiple tracers and thereby improve our knowledge of cosmological distances and the structure growth (through BAO and RSD), as well as constrain the initial conditions of the Universe and measure the sum of the mass of neutrinos. It is also possible that these surveys make observations for the 1<z<2 universe. As our study focuses on high redshift cosmology, we did  not study this possibility and left it for a future article.

In this work, we find that the BAO and RSD measurements of these future surveys will reach a sub-per-cent level but systematics may become dominant at some point. We show that the promising forecasts for NG ($\fnl\sim 1$) have to be put in perspective with the lack of knowledge of the $p$ parameter in equation \eqref{eq:pparam} which could significantly degrade these results. 
As for the neutrino mass, most of the high redshift studies will allow us to measure the mass up to a precision of 0.025 eV, even 0.01 eV when combined with CMB. Those tight constraints strongly depend on $k_\text{max}$. Different surveys provide similar constraints on $M_\nu$, and one way to significantly improve it is a better knowledge of $\tau$ and $A_s$, using independent surveys or simulations.

In addition, we choose to model an independent general survey, described by a dozen of parameters: $\{f_{\rm sky}, z_{\rm min}, \Delta z, t_{\rm exp}, p_{\rm min}, \eta_{\rm til}, S_{\rm tel}, T_{\rm sur}, N_{\rm fib}\}$, and investigate their effects on the survey accuracy, to derive the best observational strategy and the corresponding best constraints. 
We show that for NG, the survey volume should be maximised, at the expense of the tracer density. We observe a similar trend for NM, but the increase in the sky fraction between 60 per cent and 100 per cent does not imply a significant improvement in the accuracy. There has to be a density-volume trade-off. We also illustrate that for NG, the survey duration may not significantly improve the constraint\footnote{Except when the duration is very small.} and that it is better to impose a high minimum efficiency for faint objects that are crucial for mapping the high-redshift universe. 
The increase in the duration of the survey may significantly improve $\sigma(M_\nu)$. Furthermore, NM prefers a lower minimal efficiency, contrary to NG, so it is optimal to observe objects up to a higher magnitude. This model may be extended to investigate the nature of dark matter, the dark energy equation of state, and modified gravity. We aim at performing relevant studies in a follow-up paper.

\section*{Acknowledgements}
We thank Zheng Cai for providing the expected target densities for different MUST specifications, as well as comments on the manuscript.
We also thank Christophe Yeche for insightful discussions on the MSE survey properties, Noah Sailor for general discussions about spectroscopic survey forecast and 
Andreu Font Ribiera for his help with the NM forecast.\\
We acknowledge support from the Swiss National Science Foundation (SNF) “Cosmology with 3D Maps of the Universe” research grant, 200020$\_$175751 and 200020$\_$207379. Work is also supported in part by the Spanish Mineco Grant PID2019-111317GB-C32.IFAE is partially funded by the CERCA program of the Generalitat de Catalunya.
 
\section*{Data availability}
Our codes are available in GitHub on open access: \url{https://github.com/wdoumerg/Forecast_highz_spectroscopic_survey}. 

%%%%%%%%%%%%%%%%%%%%%%%%%%%%%%%%%%%%%%%%%%%%%%%%%%



%\bibliographystyle{mnras}
%\bibliography{biblio} % if your bibtex file is called example.bib
\bibliographystyle{mnras}%Used BibTeX style is unsrt
\bibliography{bibli}

%\begin{thebibliography}{99}
%\end{thebibliography}

%%%%%%%%%%%%%%%%%%%%%%%%%%%%%%%%%%%%%%%%%%%%%%%%%%

%%%%%%%%%%%%%%%%% APPENDICES %%%%%%%%%%%%%%%%%%%%%

\appendix
\section{Efficiency and SNR}
\label{sec:AppendixJiaxi}
\begin{figure}
    \centering
    \includegraphics[scale=0.55]{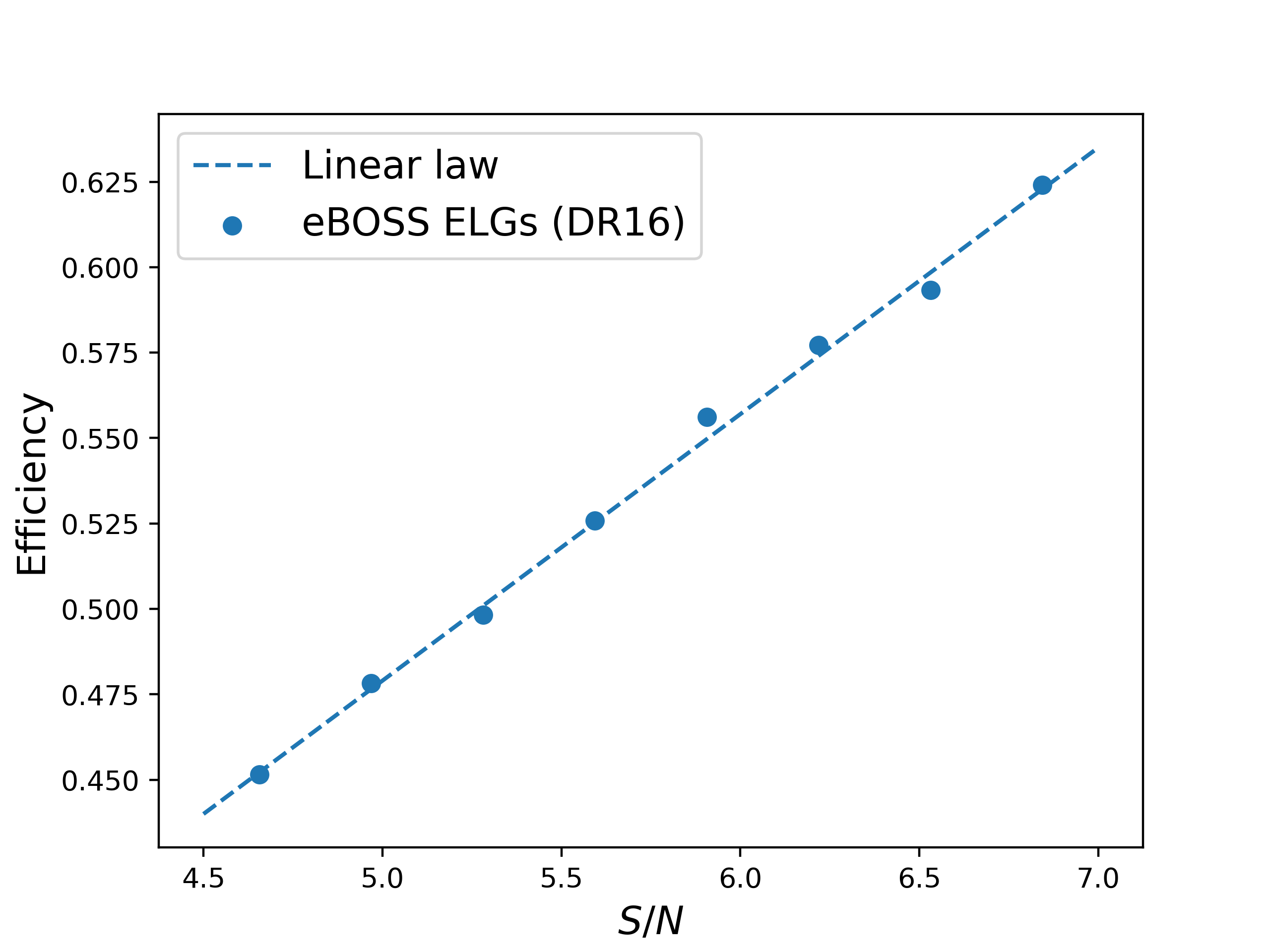}
    \caption{Efficiency as a function of the SNR, for eBOSS ELGs. We also include the linear best-fit.}
    \label{fig:jiaxi}
\end{figure}
In Figure \ref{fig:jiaxi} we present the efficiency of the ELGs observation for eBOSS DR16, as a function of the SNR for ELGs at $0.6<z<1.1$. The efficiency is the fraction of ELGs targets (photometrically defined) that are classified as ELG spectroscopically and have a reliable redshift measurement. Its efficiency scales linearly with the SNR. We assumed this trend would be similar for LBG at high redshift in our study. 

\section{Average exposure time}
\label{sec:appendixA}
Given a minimal observation efficiency, some tracers need to be observed twice or three times. Since we do not consider measurements with $m$ higher than 25 because of engineering limits, it is in practice never necessary to observe more than 3 times for a given tracer. We recall that $E(m)$ is the one-exposure efficiency law. The average number of observations is analytically given by   
\begin{equation}
    \frac{\langle t(m) \rangle}{t_\text{exp}} =\left\{
\begin{array}{ll}
1 &\text{if } (C_1) \\
1+\eta_{2}&\text{if } (C_2)\\
2-E(m)+\eta_{3}&\text{if } (C_3),
\end{array}
\right.
\end{equation}
with the ratio,
\begin{align}
    \eta_{2}&=\frac{p_\text{min}-E(m)}{\sqrt{2}E(m)} \text{ for } (C_2)\\
    \eta_{3}&=\frac{p_\text{min}-E(m)(\sqrt{2}+1-\sqrt{2}E(m))}{\sqrt{3}E(m)}\text{ for } (C_3)
\end{align}
and the conditions,
\begin{align}
    &(C_1):~E(m)\geq p_\text{min}  \text{, only one observation is necessary}\\
    &(C_2):E(m)<p_\text{min} \text{ and }E(m)(\sqrt{2}+1-\sqrt{2}E(m))\geq p_\text{min}\\
    &\hspace{0.9cm}\text{ two observations may be necessary}\\
    &(C_3): \text{higher }m\text{ than } (C_2) \text{, three observations may be necessary}.
\end{align}
Graphically, in Figure \ref{fig:new_prob}
\begin{itemize}[wide, labelwidth=!,itemindent=!,labelindent=0pt, leftmargin=0em, itemsep=.1cm, parsep=0pt]
    \item $(C1)$ represents the left part of the plane (full red), where only one observation is necessary,
    \item $(C_2)$ is the middle part, with possibly a second observation, and $\eta_2(m)$ corresponds to the the vertical width of the green area,  
    \item $(C_3)$ is the right part with possibly 3 observations, and $\eta_{3t_\text{exp}}$ is the vertical width of the blue area.
\end{itemize}  

\section{Maximal magnitude discretization} 
We have 
\begin{equation}
    \frac{N_\text{fib}\,T_\text{sur}}{4\pi\eta_{\rm til}  f_\text{sky}}=\int_{22.5}^{m_\text{max}}dm \int_{z_\text{min}}^{z_\text{min}+\Delta z} dz \underbrace{\frac{d\chi}{dz}\chi^2\langle t(m)\rangle\phi_\text{LBG}(m,z)}_{f_{mz}(m,z):=}\label{eq:A1}
\end{equation}
We define two small steps $\delta_m$ and $\delta_z$ (typically 0.01) and transform our integrals into Riemann sums. We will thus find recursively the integer $j_\text{max}$ such that the sum
\begin{equation}
    S=\sum_{j=0}^{j_\text{max}}\delta_m \sum_{i=0}^{\left\lfloor\Delta z/\delta_z\right\rfloor} \delta_z f_{mz}(m=22.5+j\delta_m,z=z_\text{min}+i\delta_z)
\end{equation}
is equal to the constant on the left-hand side of equation \eqref{eq:A1}, and deduce an approximation for $m_\text{max}$. We assume that $m_{\rm max}$ cannot be greater than 25, for practical reason since observation of fainter objects would be challenging, and because our luminosity function model does not describe LBG for higher magnitude \citep{bias_luminosity}. That is why we observed saturation  in Figure \ref{fig:density_dz_fsky} for small volume. As the luminosity function decreases with redshift, the saturation value that corresponds to observation of every galaxy up to $m_{\rm max}=25$ decreases with $\Delta z$.
\label{sec:appendixB}

\label{lastpage}
\end{document}